\documentclass[%
  superscriptaddress,
  twocolumn,
  10pt,
  aps,
  pre,
  nofootinbib, 
  final,letterpaper
]{revtex4-2}

\usepackage{graphicx}
\usepackage{hyperref}
\usepackage[utf8]{inputenc}
\usepackage{float}
\usepackage{amsmath,amsfonts,amssymb}

\usepackage{physics}

\usepackage{color}

\begin{document}

\title{Epidemic Spreading and Digital Contact Tracing: Effects of Heterogeneous Mixing and Quarantine Failures}

\author{Abbas K. Rizi}
\affiliation{Department of Computer Science, School of Science, Aalto University, FI-0007, Finland}
\author{Ali Faqeeh}
\affiliation{Department of Computer Science, School of Science, Aalto University, FI-0007, Finland}
\affiliation{Mathematics Applications Consortium for Science \& Industry, University of Limerick, Ireland}
\affiliation{CNetS, School of Informatics, Computing, \& Engineering, Indiana University, Bloomington, IN, USA}
\author{Arash Badie-Modiri}
\affiliation{Department of Computer Science, School of Science, Aalto University, FI-0007, Finland}
\author{Mikko Kivelä}
\affiliation{Department of Computer Science, School of Science, Aalto University, FI-0007, Finland}

\date{\today}

\begin{abstract}
Contact tracing via digital tracking applications installed on mobile phones is an important tool for controlling epidemic spreading. Its effectivity can be quantified by modifying the standard methodology for analyzing percolation and connectivity of contact networks. We apply this framework to networks with varying degree distributions, numbers of application users, and probabilities of quarantine failures. Further, we study structured populations with homophily and heterophily and the possibility of degree-targeted application distribution.
Our results are based on a combination of explicit simulations and mean-field analysis.  They indicate that there can be major differences in the epidemic size and epidemic probabilities which are equivalent in the normal SIR processes. Further, degree heterogeneity is seen to be especially important for the epidemic threshold but not as much for the epidemic size. The probability that tracing leads to quarantines is not as important as the application adoption rate.
Finally, both strong homophily and especially heterophily with regard to application adoption can be detrimental. Overall, epidemic dynamics are very sensitive to all of the parameter values we tested out, which makes the problem of estimating the effect of digital contact tracing an inherently multidimensional problem.
\end{abstract}

\keywords{Epidemic Spreading, percolation, contact tracing, compartmental models, covid-19} 
\maketitle

Until effective vaccines are widely deployed in a pandemic era,  carefully timed non-pharmaceutical interventions \cite{Non-pharmaceuticalPERRA2021}
such as  wearing face masks \cite{cheng2021face}, school closures, travel restrictions and contact tracing \cite{keeling2011modeling, foege1971selective, swanson2018contact, rutherford1988contact, fox2013contact} are the best tools we have for curbing the pandemic. Contact tracing is an attempt to discover and isolate asymptomatic or pre-symptomatic (exposed) individuals. In the absence of herd immunity, contact tracing is a potent low-cost intervention method since it puts people into quarantine where and when the disease spreads. Therefore, it can have a significant role in containing a pandemic by relaxing social-distancing interventions \cite{aleta2020modelling}, providing an acceptable trade-off between public health and economic objectives \cite{walker2020impact, kissler2020projecting}, developing sustainable exit strategies \cite{bell2021beyond, gilbert2020preparing}, identifying future outbreaks \cite{Koganeabd6989}, and reaching the ‘source’ of infection \cite{kojaku2021effectiveness}.

Thanks to the emergence of low-cost wearable health devices \cite{salathe2012digital, natarajan2020assessment, seshadri2020wearable, oran2020prevalence, morales2020impact,quer2021wearable, mishra2020pre, stehle2011high} and mobile software applications, digital contact tracing can now be deployed with higher precision without the problems of manual contact tracing, such as the tracing being slow and labor-intensive or people's hesitation to give identifying data about their contacts due to blame, fear, confusion, or politics. On the other hand, smartphones and wearable devices also offer continuous access to real-time physiological data, which can be used to tune other non-pharmaceutical or pharmaceutical strategies. Modern apps enable us to monitor COVID-19 symptoms \cite{ganguli2020rapid, fozouni2020direct, menni2020loss}, identify its hotspots \cite{varsavsky2021detecting}, track mosquito-borne diseases such as Malaria, Zika and Dengue \cite{ganguli2017hands, sousa2020citizen}, and detect microscopic pathogens.

In both forms---manual \cite{becker2005controlling, eichner2003case, eames2003contact, foege1971selective, swanson2018contact, kiss2006infectious, howell1997partner, fraser2004factors, kretzschmar1996modeling, klinkenberg2006effectiveness, PhysRevE.66.056115} and digital \cite{rodriguez2021population, braithwaite2020automated, salathe2020early, o2021contact, cencetti2021digital, Hambridge2021.03.09.21253198, Barrat2020.07.24.20159947}---contact tracing has been commonly considered as an effective strategy and different empirical data sets have validated this claim in short-time population-based controlled experiments \cite{salathe2020early,  rodriguez2021population}. It has been estimated that for every percentage point increase in app-users, the number of cases can be reduced by 2.3\% (in statistical analysis) \cite{wymant2021epidemiological}. However, 
such a linear view of the benefits of the app usage is likely too simplistic and ignores the complexities disease spreading, especially in heterogeneous populations \cite{pastor2007evolution, dorogovtsev2013evolution, albert2002statistical, pastor2001epidemic}. For instance, degree-heterogeneity in the contact network \cite{newman2018networks} can alter epidemiological properties in the form of variance in final outbreak size \cite{hebert2020beyond}, vanishing epidemic threshold \cite{boguna2003absence, pastor2001epidemic}, hierarchical spreading \cite{barthelemy2004velocity}, strong finite-size effects \cite{pastor2002epidemic} and universality classes for critical exponents \cite{dhara2021critical}.  Moreover, the existence of super-spreaders dictates the extent to which a virus spreads in a bursty fashion \cite{clemenccon2015statistical, lloyd2005superspreading, kupferschmidt2020some}, especially when there is high individual-level variation in the number of secondary transmissions \cite{feld1991your, barthelemy2004velocity, newman2005threshold}. Therefore, to evaluate the effectiveness of contact-tracing, degree-heterogeneity and app adoption of super-spreaders \cite{liu2020secondary, shen2004superspreading} should be taken into account. Note that in some parameter settings, contact tracing may not be effective enough \cite{aleta2020modelling, ferretti2020quantifying, sapiezynski2020fallibility}.

A potentially important factor in the effectiveness of the contact tracing apps is related to how the app-using and non-app-using populations are mixed. Several studies have shown that similar people with similar features are more likely to be in contact with each other than with people with different types of features. This phenomenon is known as homophily \cite{mcpherson2001birds,  funk2010modelling, centola2011experimental}. It has been reported in app adoption directly \cite{salathe2020early}, and indirectly through correlation in app adoption and other features exhibiting homophily, such as jobs, age, income and nationality \cite{von2021drivers, lopez2020anatomy, munzert2021tracking}. Therefore, the fraction of population that adopts the app is not the only important factor for reducing the peak and total size of the epidemic, but also the amount of homophily in app adoption can potentially have a significant role.

 Since the World Health Organization has declared the COVID-19 outbreak as a Public Health Emergency of International Concern, network scientists have developed different approaches towards analyzing epidemic tracing and mitigation with apps. Using the toolbox of network science, different groups have investigated the effectiveness of contact tracing based on the topology and directionality of contact networks \cite{Barrat2020.07.24.20159947,kojaku2021effectiveness, bianconi2021message, kryven2021contact, allard2020role, simoes2021tracking, abueg2020modeling, bassolas2021optimising, kendall2020epidemiological}. Recently, a mathematical framework aimed at understanding how homophily in health behavior shapes the dynamics of epidemics has been introduced by Burgio \textit{et al}. \cite{burgio2021impact}. This study expanded the model of Bianconi et al. \cite{bianconi2021message} and computed the reproduction number and attack rate in a homophilic population using mean-field equations.  
 
Our study investigates the effect of varying app coverage on the epidemic's threshold, probability and expected size in homogeneous and heterogeneous contact networks with and without homophily or heterophily in app adoption. Further, we explore the effect of distributing the apps randomly and preferentially to high-degree nodes \cite{bianconi2021message} in these scenarios.
 Our main focus is on the epidemic threshold and the final size of the epidemics. Therefore, we assume the dynamics of the epidemic to be governed by the simple SIR model \cite{newman2018networks}.  This model can be easily mapped to a static bond-percolation problem \cite{newman2002spread, pastor2015epidemic} so that the epidemiological properties can be measured based on the topological structure of the underlying network \cite{albert2000error, cohen2000resilience, barrat2008dynamical, dorogovtsev2008critical,allard2020role, newman2018networks}.  Note that, more complex disease transmission models, such as SEIR models in which there is an infected-but-not-contagious period E, are also covered by this formalism \cite{newman2002spread,kenah2011epidemic}.
 The difference in the spreading framework with the app to the normal one is that the infection cannot spread further if it passes a link between two app-users (app-adopters). That is, the infection process model needs to include the memory of the type of node it is coming from. We then extend the percolation framework such that we can add memory \cite{basnarkov2020random, shu2014effects} to it in order to keep track of the infection path. This leads to the observation that the epidemic size is not the same as the epidemic probability as it would be in this model without the app-users \cite{miller2007epidemic}.

Our results are largely based on mean-field-type calculations of the percolation problem, which are confirmed by explicit simulations of SIR epidemic process and measurements of component sizes in finite networks. Our findings show that: 1) the number of app-users has a direct effect on the epidemic size and epidemic probability and the difference between these two observables is larger in high-degree targeting strategy; 2) epidemics can be controlled to a much better degree in the high-degree targeting strategy; 3) even though degree-heterogeneity can strongly affect or even eliminate the epidemic threshold, high-degree targeting strategy can compensate this effect and increase the threshold significantly; 4) increasing heterophily from random mixing always increases the outbreak size and lowers the epidemic threshold; 5) increasing homophily does the opposite until an optimum, that is below the maximum homophily case, is reached; and finally 6) the probability of contact tracing succeeding in preventing further infections is not as crucial as the fraction of app-users, but can still have significant effects on the epidemic size and epidemic threshold. The only exception is when the apps are distributed to heterogeneous networks with the high-degree targeting strategy.

 \section{Modelling approach}
 
 \subsection{Disease model and connection to percolation}
 
We employ a SIR disease model on networks with additional dynamics given by the disease interactions in the presence of the disease tracking application. In the model, without the tracking application, an infected (I) node will eventually infect a neighboring susceptible (S) node with a transmission probability $p$ independently of other infections. The simulations are performed with a model where the infected nodes try to infect their susceptible neighbors with independent Poisson processes with rate $\beta$ and go to the removed state (R) after fixed time $\tau$. The fixed recovery time ensures that every infected individual, regardless of app adoption, can infect a susceptible neighbor independently with a transmission probability $p=1-e^{-\beta \tau}$ \cite{newman2002spread,kenah2007second}. These assumptions allow us to study the SIR processes using component size distributions of undirected networks where parts of the links are randomly removed \cite{grassberger1983critical,newman2002spread, kenah2007second, kenah2011epidemic, miller2007epidemic}: an epidemic starting from a single node can reach any other node exactly when there is a path of such transmitting links connecting them, i.e., they are in the same component in a network where the potential contact links are removed with probability $p$.
Thus, the epidemic threshold, epidemic probability and epidemic size can be read from percolation simulations \cite{grassberger1983critical,newman2002spread, kenah2007second, kenah2011epidemic, miller2007epidemic} (see Section~\ref{sec:components}).
Note that without fixed recovery time, the presence of spreading paths through neighboring links would not be independent, and this would not be a bond percolation problem in an undirected network where edges are removed independently. However, the epidemic threshold, final epidemic size, and the expected outbreak size below the epidemic threshold would still be correctly predicted by this methodology \cite{kenah2007second,miller2007epidemic}.

We model the effect of applications to the disease spreading as follows: if an app-user infects another app-user, that second node will get infected but will quarantine themselves with probability $p_{\rm app}$. The quarantined user will have no further connections that would spread the infection they received from the other app-user. A substantial deviation from a realistic spreading case in our model is that the quarantine does not prevent the disease spreading to the quarantined node through a third node.
That is, we only model the primary infection path from the other app-user causing the alarm but do not stop the possible concurrent secondary infection paths from a third node. 
Strictly speaking, this simplification in the modeling returns a lower bound on the effectiveness of the app-based contact tracing, but given that our contact network models are sparse random graphs (see Section~\ref{sec:models}) that do not contain local loops, the difference can only be observed if a large enough fraction of the population is infected at the same time. Critically, this does not affect the epidemic threshold but could have implications for parameter regions where the epidemic size is large, depending on the quarantine durations.

\begin{figure}[!htb]
\includegraphics[width=1\linewidth]{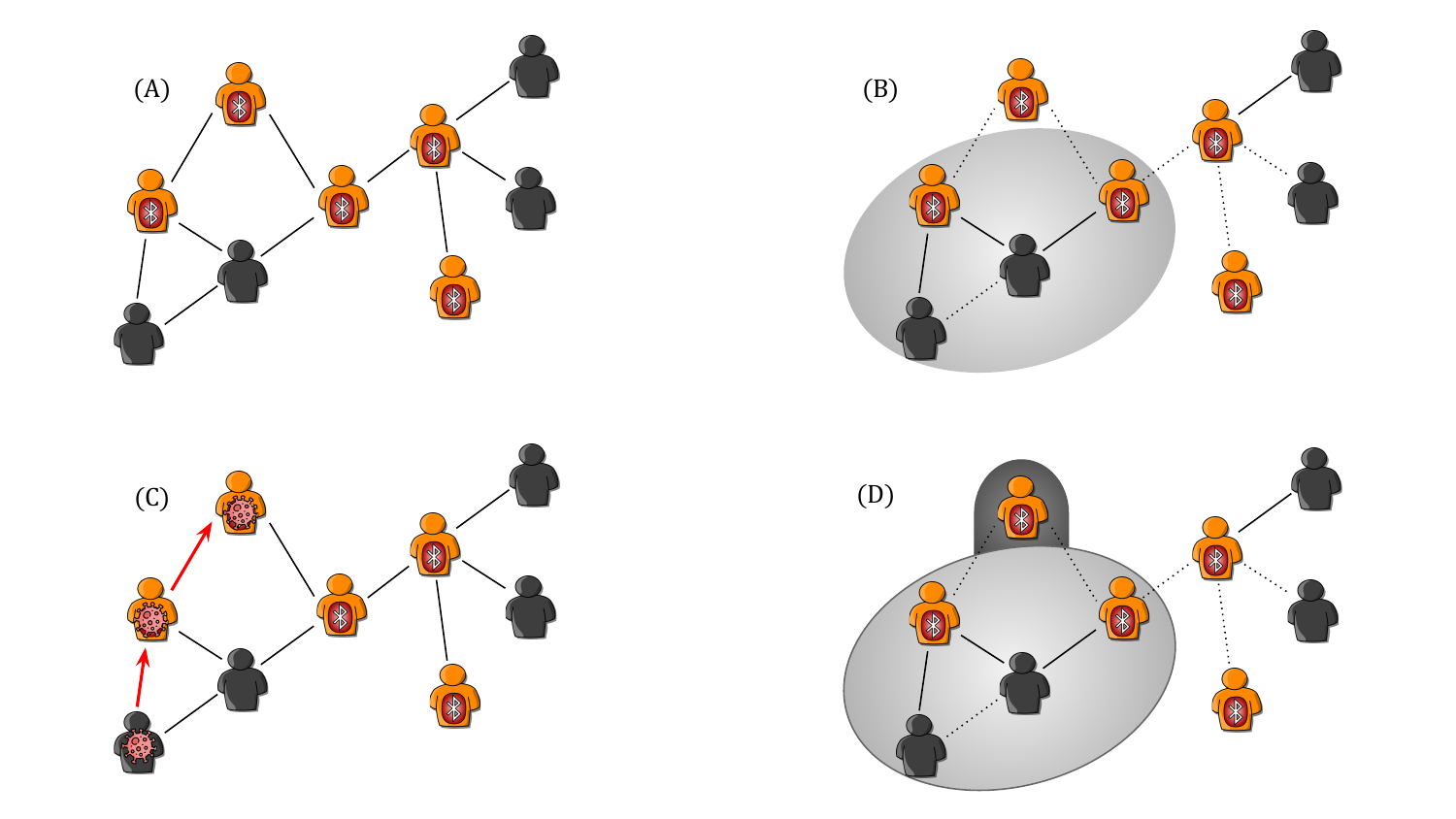}
\caption{(A) Original contact network with app-users marked with the oval symbol. (B) The normal largest component, after the dotted links have been removed in the percolation process at random. When apps are working perfectly, links between a pair of app-users are removed with probability $p_{\rm app} = 1$ and other links are removed with probability $p$. (C) An example of a path of infection: the second app-user can be infected; therefore, it must be included in the outbreak size. (D) Extending the giant component to include the secondary app infections. The second infected app-adopter is added to the giant component with transmission probability $p$.}
\label{fig:extension}
\end{figure}

The SIR spreading process can be mapped to a slightly more complicated percolation problem in the presence of apps \cite{bianconi2021message,Barrat2020.07.24.20159947}.
To model app-user quarantines, one needs to delete the links between two app-users with the probability of successful quarantine due to the app, $p_{\rm app}$. This ensures that we ignore the infection paths through two app-users when one of them is successfully quarantined. However, removing these links also removes the second app-user from the component, even though they are infected. To correct this, we need first to find the network components and then extend them by including all app-users outside of the component connected to another app-user (and considering the probability $p$ that the link is kept). See Fig.~\ref{fig:extension} for an illustration of this process, which leads us to two definitions of components: normal and extended.

\subsection{Components, epidemic size and epidemic probability}
\label{sec:components}

In the SIR model without apps, the component size distribution can be used to describe the late stages of the epidemics approximately. Given an initially infected node, the size of the component it belongs to determines the size of the outbreak.
In an infinitely large population, we say that an outbreak is an epidemic if it spans a non-zero fraction of the population.
The relationship between percolation and the final epidemic size is straightforward if the population is large enough that it can be approximated with an infinite undirected transmission network \cite{newman2002spread,miller2007epidemic}. In this case, the percolation threshold gives the epidemic threshold and below it, an outbreak always spans only a zero fraction of the population because all the components are of finite size. Above the percolation threshold, there is a single giant component that spans $s_{\max} = S_{\max}/N$ fraction of the nodes. This is equivalent to both the size of the epidemic, given that there is one, and the probability that there is an epidemic starting from a single initially infected node \cite{newman2002spread,miller2007epidemic}; $s_{\max}$ is the fraction of nodes that can be reached from the giant component (epidemic size) and the probability that randomly chosen node belongs to the giant component (probability of the epidemic).
The expected epidemics size in a fraction of eventually infected nodes is, in this case, given as $s^2_{\max}$. 

When we introduce apps to the spreading process, the equivalence of the epidemic size and epidemic probability breaks down.
Both the normal component and the extended component become important. The component size still gives the probability that there is an epidemic, as is the case without the apps. However, the epidemic size, given that there is one, is now given by the extended component size $s'_{\max}$. The expected epidemic size is then given by $s_{\max} s'_{\max}$.

Similar relationships hold for finite-size systems.
For example, the expected size of the epidemics from single source becomes
\begin{eqnarray}
\label{eq:expected_size}
    \langle E \rangle = \sum_c \frac{S_{c}}{N} S'_{c} \,,
\end{eqnarray}
where $S_{c}$ is the normal size and $S'_c$ is the extended size of the component $c$ and $N$ is the total number of nodes. In this formula, $S_{c}/N$ gives the probability that the initially infected node is in the component $c$ and $S'_c$ gives the size of the epidemic if a node in component $c$ is chosen.

\subsection{Network models}
\label{sec:models}

We aim to study how the network topology, amount and distribution of app-users over the network affect the epidemics. We study networks with degree distribution $P(k)$ and average degree $\left \langle k \right \rangle$ such that each node is an app-user with probability $\pi_{\rm a}$ and not an app-user with probability $1-\pi_{\rm a}$. We distribute the app-users with one of two strategies: 1) uniformly at random or 2) by distributing the apps in the order of their degree such that the high-degree nodes get the apps first. 

We use three different models to generate the network topology.
We use i) Poisson ($\mathrm{ER}$) random graphs \cite{bollobas2001random} to model homogeneous contact patterns and ii) scale-free networks generated with the Chung–Lu model (CL) \cite{chung2002average, chung2002connected} to model heterogeneous networks. In generation of CL networks, the expected degree of each node is drawn from a continuous power-law distribution $P(k)\propto k^{-3}$ such that the minimum expected degree is set to a value that gives us the expected average degree $\left \langle k \right \rangle$ of our choice. Given a sequence of expected degrees $W = \{w_1, w_2, ..., w_n\}$, Miller algorithm \cite{miller2011efficient} assigns a link between node $u$ and node $v$ with probability $p_{uv}\propto w_uw_v$. This algorithm returns a network without multiple links with almost the same power-law degree distribution. 

We model homophily (and heterophily) with regards to apps usage with
iii) a modular network model (MN) introduced in Ref~\cite{Gleeson08, melnik2014dynamics} with two groups of nodes: app-users and non-app-users. This model starts with a degree sequence produced either by the $\mathrm{ER}$ or CL models and connects the nodes depending on which groups they belong with probabilities $\pi_{\rm aa}$ (app-user to app-user), $\pi_{\rm an}$ (app-user to non-app-user), $\pi_{\rm na}$ (non-app-user to app user), and $\pi_{\rm nn}$ (non-app-user to non-app-user).
We only need to fix one of these probabilities, $\pi_{\rm aa}$, and other types of links are formed with probabilities 
$\pi_{\rm an} = 1 - \pi_{\rm aa}$, $\pi_{\rm na} = \frac{\pi_{\rm a}}{1-\pi_{\rm a}}(1- \pi_{\rm aa})$ and $\pi_{\rm nn}=1-\pi_{\rm na} = \frac{1-\pi_{\rm a} - \pi_{\rm a}(1-\pi_{\rm aa})}{1-\pi_{\rm a}}$, where $\pi_{\rm a}$ is the probability that a person is an app-user and the second equality comes from the balance between the number of links from app-users to non-app-users and from non-app to app-users, that is, $\pi_{\rm a}N \pi_{\rm an} \langle k\rangle  =  (1-\pi_{\rm a})N\pi_{\rm na}\langle k\rangle $. The numerical simulations of the MN work by randomly choosing a group for half edges with the given probabilities and matching them to each other uniformly randomly. This can lead to self-links and multi-links, which these are discarded after the randomization procedure.

While there is no correlation between the app adoption status in homogeneous (ER) or heterogeneous (CL) networks above, in the third model (MN), the existence of homophily or heterophily of the network structure is determined by comparing $\pi_{\rm aa}$ to its value for the neutral case with no homophily or heterophily. In the absence of homophily or heterophily, $\pi_{\rm aa} = \eta_{\rm a}$, where $\eta_{\rm a}$ is the ratio of the number of links that emerge from app-users to the total degree; this is because if the nodes were connected purely at random, the probability that a link from an app-user connects it to another app-user equals the ratio of the number of stubs that app-users have to the total number of stubs, i.e., $\eta_{\rm a}$.
In the case of a random selection of app-users $\eta_{\rm a}=\pi_{\rm a}$, since both app-users and non-app-users have on average the same number of stubs and the fraction of stubs that app-users have equals the fraction of app-users in the system, i.e., $\pi_{\rm a}$.
In a high-degree targeting strategy, the number of stubs that app-users have on average is larger than that of non-app-users. In that case, $\eta_{\rm a}$ can be calculated from the degree distribution (see Sec.~\ref{sec:Ali}).
When $\pi_{\rm aa} > \eta_{\rm a}$, app-users are more likely to be connected to each other than in a network in which a fraction of $\eta_{\rm a}$ of them being uniformly randomly placed. On the other hand, when $\pi_{\rm aa} < \eta_{\rm a}$ nodes are more likely to be connected to the nodes of the other type (heterophilic network). In the heterophilic regime, for some pairs of $(\pi_{\rm aa},  \pi_{\rm a})$, networks are not realizable because of the constraints explained in Sec.~\ref{sec:Ali}. The white region in Fig.~\ref{fig:4_branching} shows the region of $\pi_{\rm aa}$\nobreakdash-$\pi_{\rm a}$ plane that networks cannot be created in that parameter space.

\section{Analytic and simulation methods} \label{sec:theory}

The epidemics are studied here with various methods of approximation. We employ analytical computations based on mean-field-type approximations to efficiently analyze our models' wide parameter space and provide explicit formulas for our main observable quantities. Here an approximation based on branching processes \cite{vazquez2021multitype} can be used to determine the critical point. Following Ref.~\cite{Barrat2020.07.24.20159947}, a more detailed calculation based on percolation arguments will give us the component sizes which can be related to the final epidemic size and epidemic probability. Simulations of the network connectivity then complement these mean-field approximations. Finally, we explore the accuracy of the mean-field approximations via explicit simulations of the SIR model.

\subsection{Giant component size from consistency equations}\label{sec:Ali}

To study the behavior of the epidemic dynamics, we form consistency equations for the giant component size. In Ref.~\cite{Barrat2020.07.24.20159947} the governing equations for the size of the epidemic and the transition point were obtained for the case of random networks in the absence of homophily.
Here we derive the analytical results for the more general case of the spectrum of heterophilic to homophilic networks, a special case of which is the non-homophilic networks of Ref.~\cite{Barrat2020.07.24.20159947}.
We consider that app-users and non-app-users might be connected together with a pattern different from pure random chance using the modular network model (MN).

We aim to write the self-consistent equations for the probability, $u_{\rm n}$, that following a link to a non-app-user does not lead to the giant component and probability $u_{\rm a}$, that following a link to an app-user does not lead to the giant component. Using these probabilities, the relative size of the giant component $s$ and the relative size of the extended giant component $s'$ can be obtained, where $s$ is, in fact, the fraction of nodes infected through non-app-users, while $s'$ also includes individuals who caught the infection through an app-user before they could quarantine themselves (see Sec.~\ref{sec.extension}).

We need to know the probability $u_{\rm n}$ ($u_{\rm a}$), that a randomly chosen link leading to a non-app-user (app-user) is not in the giant component. The probability that a non-app-user (app-user) is not connected to the giant component via a particular neighboring node is equal to the probability that \textit{that} non-app-user (app-user) is not connected to the giant component via any of its other neighbors. A non-app-user is connected to another non-app-user with probability $\pi_{\rm {nn}} = 1-\pi_{\rm {na}}$ and to an app-user with probability $\pi_{\rm {na}}$.
So, a link leading out from a
non-app-user does not lead to the giant component if it leads to another non-app-user that is not in the giant component (which happens with probability $(1-\pi_{\rm na})u_{\rm n}$)
or an app-user that is not in the giant component (which happens with probability $\pi_{\rm na}u_{\rm a}$). That is, the total probability for following a link out from a non-app-user not leading to the giant component is $u_{n \rightarrow} = (1-\pi_{\rm na})u_{\rm n}+\pi_{\rm na}u_{\rm a}$. Since the degree of neighboring nodes is disturbed according to the excess degree distribution $q_k$, the probability that a non-app-user that is encountered by following a link to it is not connected to 
the giant component via any of its $k$ neighbors is $\sum_k q_k u_{n \rightarrow}^k$. This probability is, by definition, $u_{\rm n}$, leading to the self-consistent equation below for $u_{\rm n}$:
\begin{equation}
\label{eq:gfM0}
    u_{\rm n} = g_1( (1-\pi_{\rm na})u_{\rm n}+\pi_{\rm na}u_{\rm a}),
\end{equation}
where $g_1$ is the generating function for excess degree distribution \cite{newman2018networks}. To find $u_{\rm a}$, we can use the same treatment, except that we should consider how app-app connections depend on the probability of success in contact tracing \cite{Barrat2020.07.24.20159947}. If $p_{\rm app}$ is the probability the apps work as expected, then $1-p_{\rm app}$ is the probability that the app-user does not effectively quarantine after being been in contact by an infectious app-user. Therefore, $u_{\rm a}$ can be expressed as the self-consistent equation below: 
\begin{equation}
    u_{\rm a} = g_1\big((1-\pi_{\rm aa})u_{\rm n}+\pi_{\rm aa}(p_{\rm app}+(1-p_{\rm app})u_{\rm a})\big).
    \label{eq:gfMa}
\end{equation}

Note that $\pi_{\rm na}$ is determined by the free parameters $\pi_{\rm a}$ and $\pi_{\rm aa}$ as we already showed that $\pi_{\rm na}=\frac{\pi_{\rm a}}{1-\pi_{\rm a}} (1-\pi_{\rm aa})$.

Given $u_{\rm n}$ and $u_{\rm a}$, the average probability that a node belongs to the giant component, or equivalently the
fraction of the network occupied by the giant component, is now given by:
\begin{eqnarray}
 \label{eq:normal_comp_size_theory}
    s = 1  - (1-&&\pi_{\rm a}) g_0\big( (1-\pi_{\rm na})u_{\rm n}+\pi_{\rm na}u_{\rm a}\big) \nonumber\\
    -&& \pi_{\rm a} g_0\big((1-\pi_{\rm aa})u_{\rm n}\nonumber\\
 &~&~~~~~~~+
 \pi_{\rm aa}(p_{\rm app}+(1-p_{\rm app})u_{\rm a})\big)
   \,,
\end{eqnarray}
where $g_0$ is the generating function for degree distribution. We can approximate $s'$ by writing:
\begin{eqnarray}
 \label{eq:ex_comp_size_theory}
    s' = 1  - (1-&&\pi_{\rm a}) g_0\big( (1-\pi_{\rm na})u_{\rm n}+\pi_{\rm na}u_{\rm a}\big) \nonumber\\
    -&~& \pi_{\rm a} g_0\big((1-\pi_{\rm aa})u_{\rm n} 
    + \pi_{\rm aa}u_{\rm a})   \big)
   \,,
\end{eqnarray}
where, as opposed to Eq.~\ref{eq:normal_comp_size_theory}, the third term is not a function of $p_{\rm app}$ and the reason is that  Eq.~\ref{eq:normal_comp_size_theory} assumes that if the app works (which happens with probability $p_{\rm app}$) then the probability that a link connected to an app-user does not lead to the giant component is 1 (while if the app does not work it is $u_{\rm a}$). However, whether the app works or not, the probability that an app-user does not get infected from another app-user is $u_{\rm a}$. When apps work, if the second app-user is infected, she quarantines herself and does not infect any other node).

In the case of including a transmission probability $p$ which is less than 1 (in the above equations it was assumed the links are transmitting with probability 1), Eqs.~\ref{eq:gfM0} and \ref{eq:gfMa} will change to:
\begin{eqnarray}
 u_{\rm n} =  1-p + pg_1&(& (1-\pi_{\rm na})u_{\rm n}+\pi_{\rm na}u_{\rm a}) \,,
 \\
 u_{\rm a} = 1-p + p
 g_1 &(&
 (1-\pi_{\rm aa})u_{\rm n}\nonumber\\
 &~&~+
 \pi_{\rm aa}(p_{\rm app}+(1-p_{\rm app})u_{\rm a})
 )\,.
\end{eqnarray}

When the fraction $\pi_{\rm a}$ of nodes selected to adopt the app are all the highest degree nodes in
the network, these nodes all have a degree higher than $k_{\rm a}-1$ such
that they include some of $k_{\rm a}$ nodes and the rest are comprised of all nodes with degree larger
than $k_{\rm a}$. Then 
for the fraction $\eta_{\rm a}$ of the links protruding from the app-users (which are
the top $\pi_{\rm a}$ fraction of nodes) we can write:
\begin{eqnarray}
 \eta_{\rm a}&= &   r^*k_{\rm a}p_{k_{\rm a}}/\langle k\rangle  +\sum_{k_{\rm a}+1}^{\infty}                     k p_k/\langle k\rangle \,\\
                     &=&             \sum_{k_{\rm a,right}}^{\infty} k p_k/\langle k\rangle
                                            \,, \label{eq:eta1}
\end{eqnarray}
where $r^*$ is the fraction of degree $k_{\rm a}$ nodes that are app-users and in Eq.~\ref{eq:eta1}
we absorbed $r^*$ into $p_k$ so that $p_{k_{\rm a,right}} = r^*p_{k_{\rm a}}$ represents
the fraction of nodes in the network that have degree $k_{\rm a}$ and are app-users (so in Eq.~\ref{eq:eta1},
$k_{\rm a,right}$ takes the value $k_{\rm a}$).

Then for a network with homo/heterophily:
\begin{eqnarray}
 u_{\rm n} &=& 1-p + \frac{p}{1-\eta_{\rm a}} \sum_{k=0}^{k_{\rm a,left}}
    q_k[ (1-\pi_{\rm na})u_{\rm n}\nonumber\\
    &~&~~~~~~~~~~~~~~~~~~~~~~~~~~~~~~~~~~~+
    \pi_{\rm na}u_{\rm a}]^k \,,\\
  u_{\rm a} &=& 1-p + \frac{p}{\eta_{\rm a}}\sum_{k_{\rm a,right}}^{\infty}
    q_k[
      (1-\pi_{\rm aa})u_{\rm n}\nonumber\\
 &~&~~~~~~~~~~~~~~~~+
 \pi_{\rm aa}(p_{\rm app}+(1-p_{\rm app})u_{\rm a}) ]^k\,, 
\end{eqnarray}
and
\begin{eqnarray}
 s =  1 &-& \sum_{k=0}^{k_{\rm a,left}}p_k\left[ (1-\pi_{\rm na})u_{\rm n}+\pi_{\rm na}u_{\rm a}\right]^k \nonumber\\
     &-&\sum_{k_{\rm a,right}}^{\infty}p_k[  (1-\pi_{\rm aa})u_{\rm n}\nonumber\\
 &~&~~~~~~~~~~~+
 \pi_{\rm aa}(p_{\rm app}+(1-p_{\rm app})u_{\rm a})]^k
           \,. 
\end{eqnarray}
A special case of which are networks with neutral (non-existing) homophily, where $\pi_{\rm aa}$
is obtained to be equal to $\eta_{\rm a}$ and accordingly $\pi_{\rm na} = \eta_{\rm a}$, therefore,
\begin{eqnarray}
 u_{\rm n} &=& 1-p + p\frac{1}{1-\eta_{\rm a}} \sum_{k=0}^{k_{\rm a,left}}
      q_k\left[ (1-\eta_{\rm a})u_{\rm n}+\eta_{\rm a}u_{\rm a}\right]^k \,,\\
 u_{\rm a} &=& 1-p + p\frac{1}{\eta_{\rm a}}\sum_{k_{\rm a,right}}^{\infty}
   q_k[~ \eta_{\rm a}(p_{\rm app}+(1-p_{\rm app})u_{\rm a})
   \nonumber\\
     &~&~~~~~~~~~~~~~~~~~~~~~~~~~~~+
     (1-\eta_{\rm a})u_{\rm n} ~]^k\,, 
\end{eqnarray}
and
\begin{eqnarray}
 s =  1&-& \sum_{k=0}^{k_{\rm a,left}} p_k\left[ (1-\eta_{\rm a})u_{\rm n}+\eta_{\rm a}u_{\rm a}\right]^k\nonumber\\
                &-& \sum_{k_{\rm a,right}}^{\infty} p_k[\eta_{\rm a}(p_{\rm app}+(1-p_{\rm app})u_{\rm a})
   \nonumber\\
     &~&~~~~~~~~~~~~~~+
     (1-\eta_{\rm a})u_{\rm n}]^k
                \,. 
\end{eqnarray}
These results predict the behavior of the epidemic dynamics in the thermodynamic limit. Therefore they describe the dynamics very well when the network size is large enough.

\subsection{Mean-field approximation for the branching process} \label{sec:branching}

An alternative to writing the consistency equations for the giant component size is to assume that a branching process governs the epidemic dynamics. Then, a straightforward way of finding the epidemic threshold in the SIR model is to find the critical point of a branching process, where the branching factor is given by the expected excess degree  $q$. In the epidemic setting, the branching factor $\bar{k}_{\rm e}= p  q $ gives the expected number of people one infected person infects during the epidemic process. Note that the branching factor has been used as the definition of the basic reproduction number $R_0$ \cite{miller2007epidemic}, but 
is different from the basic reproduction number when it is defined in the networks as $R_0=\beta/\gamma \langle k \rangle$ \cite{pastor2015epidemic}.
In the SIR model with the app, we need to duplicate the populations so that we separately track the ones without the app ($S_{\rm n}$, $I_{\rm n}$ and $R_{\rm n}$) and with the app ($S_{\rm a}$, $I_{\rm a}$ and $R_{\rm a}$).

 Given that the apps are uniformly distributed to $\pi_{\rm a}$ fraction of the nodes and $\bar{k}_{\rm e}$ is the branching factor, we can write a mean-field approximation based on the branching process as follows:
\begin{eqnarray}
    I_{\rm n}^{(t+1)} = \bar{k}_{\rm e} \Big( \pi_{\rm nn}I_{\rm n}^{(t)} +\pi_{\rm an} I_{\rm a}^{(t)} \Big) \,,
     \\
     I_{\rm a}^{(t+1)} = \bar{k}_{\rm e} \Big( \pi_{\rm na}I_{\rm n}^{(t)} +\pi_{\rm aa} (1-p_{\rm app})I_{\rm a}^{(t)} \Big)\,.
\end{eqnarray}

By defining $a = \pi_{\rm nn}\bar{k}_{\rm e}$, $b =  \pi_{\rm an}\bar{k}_{\rm e}$, $c =  \pi_{\rm na}\bar{k}_{\rm e}$ and $d = \pi_{\rm aa}\bar{k}_{\rm e}(1-p_{\rm app})$, the difference equations can be written as:
 \begin{equation}\label{matrix_diff}
    \mathbf{X}_{t+1} = \mathbf{AX}_{t}\,,
\end{equation}
 where $\mathbf{X}_{t} = \begin{pmatrix} I_{\rm n}^{(t)}\\ I_{\rm a}^{(t)} \end{pmatrix}$ and $\mathbf{A} = \begin{pmatrix} a & b\\ c & d \end{pmatrix}$. 
 
 The steady state $\mathbf{X}_{t+1} = \mathbf{X}_{t}$ is possible if all the eigenvalues $\lambda$ of the transition matrix $\mathbf{A}$  (whether real or complex) have an absolute value which is less than 1; 
 
\begin{equation}\label{eigen_values}
    \lambda_{\pm} = \frac{a+d}{2} \pm \sqrt{{(\frac{a+d}{2})^2 - (ad-bc)}}.
\end{equation}

Without contact tracing, there is a chance of epidemic, given the initial reproductive number is $\bar{k}_{\rm e} > 1$. In the case of app adoption, the critical value of app-users $\pi_{\rm a}^{\rm c}$ that is needed for reducing the reproductive number can be derived by setting $\lambda = 1$ which leads to:

\begin{equation} \label{eq1long}
\begin{split}
& \frac{1-\pi_{\rm a}(2 -\pi_{\rm aa})}{1-\pi_{\rm a}}\Big(\bar{k}_{\rm e} + \bar{k}_{\rm e}^2 \pi_{\rm aa} (1-p_{\rm app})\Big) \\
 & + \bar{k}_{\rm e}\pi_{\rm aa}(1-p_{\rm app}) + \frac{\bar{k}_{\rm e}^2\pi_{\rm a}(1-\pi_{\rm aa})^2}{1-\pi_{\rm a}} = 1.
\end{split}
\end{equation}

When apps work perfectly, the epidemic threshold is given by:
\begin{equation} \label{kc_homo}
\begin{split}
\bar{k}_{\rm c} = \frac{\sqrt{1+\pi_{\rm a}\pi_{\rm aa}\big[4(\pi_{\rm a}+\pi_{\rm aa})-3(\pi_{\rm a}\pi_{\rm aa}+2)\big]} }{2\pi_{\rm a}(\pi_{\rm aa} - 1)^2}&\\
+ \frac{ 2\pi_{\rm a}-\pi_{\rm a}\pi_{\rm aa}- 1}{2\pi_{\rm a}(\pi_{\rm aa} - 1)^2}&.
\end{split}
\end{equation}
For each value of $\pi_{\rm a}$ there is a non-trivial optimum value $\pi_{\rm aa}^{\rm opt}$ that leads to the largest epidemic threshold in terms of the branching factor, which is:
\begin{equation}\label{opt_po}
    \pi_{\rm aa}^{\rm opt} = \frac{\pi_{\rm a}-2}{3\pi_{\rm a}-4}.
\end{equation}

The critical app adoption can be also calculated as:
\begin{equation}\label{abbas_pc}
    \pi_{\rm a}^c = \frac{1-\bar{k}_{\rm e}}{\bar{k}_{\rm e}^2(\pi_{\rm aa} -1)^2 + \bar{k}_{\rm e} (\pi_{\rm aa} -2) + 1}.
\end{equation}

In the absence of homo/heterophily, $\pi_{\rm aa} = \pi_{\rm a}$,  Eq.~\ref{eq1long}, gives the same result as of Ref.~\cite{Barrat2020.07.24.20159947}, such that:
\begin{equation}
    \pi_{\rm a}^c =\frac{\bar{k}_{\rm e} -1 + \sqrt{\big(\bar{k}_{\rm e} - 1 \big)\big(\bar{k}_{\rm e} + 3\big)} }{2\bar{k}_{\rm e}}\,.
    \label{equ:pi_c}
\end{equation}

Vazquez \cite{vazquez2021multitype} also provides a clear way of combining different intervention strategies and shows how our specific results about application homophily are affected by other interventions.

\subsection{Component size simulations}\label{sec:Abbas}

Next, we describe how to extract the giant component in simulated networks and how these simulation results can be used to find the critical points of the disease spreading process. The component sizes can also be used to find the epidemic size distributions as described in Section~\ref{sec:components}.

\subsubsection{Component Extension}\label{sec.extension}

In each simulation run, we simulate one network structure $G$ and distribute the apps to the nodes according to one of the models described in Section~\ref{sec:models}.
From the original network $G$, we keep each link with probability $ p = 1 -e^{-\beta\tau}$, 
which is the probability of infection going through a link without apps. We also remove all the links between two app-users with probability $p_{\rm app}$ and call the resulting network $G_{\rm a}$. The components of graph $G_{\rm a}$ are the normal components.

The extended components can be reached by going through every normal component and extending it.
For every app-user $\alpha$ in the component $C$, we go through the neighbors $n_\alpha = \{ \alpha_1, \alpha_2, , ..., \alpha_{ k} \}$ in the original network $G$. If $\alpha_i$ is an app-user and not in the component $\alpha_{ i} \notin C$, we add it to the component extension $C'$ with probability $p$.
The total set of infected nodes, if starting from a node in $C$, will be $C \cup C'$.
As these are disjoint sets, we can compute the size as $S'_{C}=|C|+|C'|$ and $S_{\rm c}=|C|$.

\subsubsection{Susceptibility}
\label{Susceptibility_formula}

In numerical simulations of finite size systems, we can use the peak of a susceptibility measure to find the critical transition point. Theoretically, susceptibility \cite{dorogovtsev2008critical} is a measure of fluctuation in the component sizes, which is singular at the epidemic threshold (the critical point). In network percolation studies, it is defined as the expected growth in the size of the giant component when a random link is added to the network. Therefore, susceptibility in an ordinary percolation problem can be written as:
\begin{equation}\label{Susceptibility1}
	\chi = \frac{\sum_{c \neq c_{\max}} {S^2_c} - {S^2_{c_{\max}} } }{N - S_{c_{\max}}}\,,
\end{equation}
where $S_{c}$ is the size of the component $c$, $c_{\max} = \rm{argmax}_c S_c$ is the largest component. 

 Here, we are dealing with two types of components, and as is shown in Fig~\ref{fig:1}D, the fraction of the sum of component sizes and network size $S_{\sum}/N$ can be larger than one. Susceptibility should be a monotonically decreasing function in the supercritical regime. However, plugging the extended component sizes into Eq.~\ref{Susceptibility1} results in a growth in the tail of susceptibility, turning it to a non-monotonic function in the supercritical regime. Therefore, this formulation of susceptibility is not suitable in the current case since the maximum of Eq.~\ref{Susceptibility1} could lead to estimates of critical points that are very far from the actual one.
Instead, we can use the expected growth in the extended giant component, which can be computed as:
\begin{equation}\label{Susceptibility2}
    \chi' = \frac{\sum_{c \neq c_{\max}} S_{c} S_{c}'(1-\frac{S'_{c_{\max}}}{N})}{N - S_{c_{\max}}}\,,
\end{equation}
where $S_{ c}$ and $S'_c$ are the size and the extended size of the component $c$ and $c_{\max} = \rm{argmax}_c S'_c$ is the largest component measured in the extended size.

\subsection{Explicit compartment model simulations}
\label{sub:arashSIR}

Finally, we will perform explicit simulations of the spreading processes to confirm the theoretical results we arrived at via the approximations we presented above. The effect of tracking applications can be integrated into compartment model simulation by introducing separate {susceptible} and {infected} compartments for people with and without the app. The interactions between people with no app installed is similar to those of the normal SIR process, namely, {susceptible} individuals with no app  $(S_{\rm n})$ can become {infected} $(I_{\rm n})$ by being in contact with {infected} people that either do not have the app installed $(I_{\rm n})$ or have it installed $(I_{\rm a})$. However, if a {susceptible} individual with the app $(S_{\rm a})$ comes into contact with an {infected} individual with app $(I_{\rm a})$, they will become {infected} but they will also receive infection notification from the app which means they will be {quarantined} $(I_{\rm q})$. Quarantined individuals cannot infect anyone else. Eventually, all the infected individuals will move to the recovered compartment after a constant predetermined amount of time ($1/\gamma$) has passed from the beginning of their infection. The recovered compartment is divided into three compartments $R_{\rm n}$, $R_{\rm a}$, and $R_{\rm q}$ to track which infected compartment the node is originating from.

The set of all reactions can be written as follows:
\begin{equation}
\begin{aligned}
    &S_{\rm n} + I_{\rm n} \xrightarrow{\beta} I_{\rm n} + I_{\rm n}, \, &&S_{\rm a} + I_{\rm n} \xrightarrow{\beta} I_{\rm a} + I_{\rm n},\\
    &S_{\rm n} + I_{\rm a} \xrightarrow{\beta} I_{\rm n} + I_{\rm a}, \, &&S_{\rm a} + I_{\rm a} \xrightarrow{\beta} I_{\rm q} + I_{\rm a},\\
    &I_{\rm n} \xrightarrow{\gamma} R_{\rm n}, \, &&I_{\rm a} \xrightarrow{\gamma} R_{\rm a},\\
    &I_{\rm q} \xrightarrow{\gamma} R_{\rm q}\,.
\end{aligned}
\end{equation}
Note that while edge reactions are governed by Poisson processes happening at a constant rate $\beta$, unlike most common SIR models, node reactions are governed by constant cutoff time $1/\gamma$ and happen exactly $1/\gamma$ units of time after the infection of the node.

As interactions in the simulation are bound to take place over edges of a static network, with nodes belonging to each of the compartments, as shown in Sec.~\ref{sec:Results}, the results are similar to a component size simulation (which are described in Sec. \ref{sec:Abbas}) on a network with effective connectivity of $\bar{k}_{\rm e} = \langle k \rangle (1-e^{-\beta/\gamma})$. As only the ratio between $\beta$ and $\gamma$ plays as a parameter in the model, we set the value of $\gamma$ to 1.

In each simulation, starting from a single infected node and running the simulation in discrete time steps of $10^{-4}$ units until no further reaction is possible, the final number of nodes that end up in $R_{\rm q}$, $R_{\rm a}$ and $R_{\rm n}$ determine total size of infection corresponding to the extended component size $S'$ of the component that the initial seed node belongs to. The final combined size of the $R_{\rm n}$ and $R_{\rm a}$ component, however, represents the size of the component $S_{\rm n}$ that the seed node (index case) would belong to, had we removed app-app links. By adding $I_{\rm a}$ and $I_{\rm q}$ compartments, as compared to normal SIR processes, and linking them to the state of the source of infection and the internal state of each node, we include information about the history of the spreading agent more than one step back in the simulation of the spreading process.

\section{Numerical results}\label{sec:Results}
 
We will next illustrate using the theory and simulation introduced in Sec.~\ref{sec:theory} how the various parameters affect the epidemic sizes and epidemic probabilities. The simulation studies are done in networks of $10^4$ nodes and averaged over $10$ realizations. We use two network topologies: homogeneous networks (Erd\H{o}s-R\'enyi networks) with expected degree $\langle k \rangle = 10$ and random networks with expected degree sequence driven from power-law degree distribution $p(k) \propto k^{-3}$, with a minimum degree cutoff adjusted such that the average degree is set to $10$ \cite{miller2011efficient}.

 \subsection{Differences in normal and extended components}\label{sec:ResultsGIANT}
 
 \begin{figure}[!htb]
 \includegraphics[width=1.\linewidth]{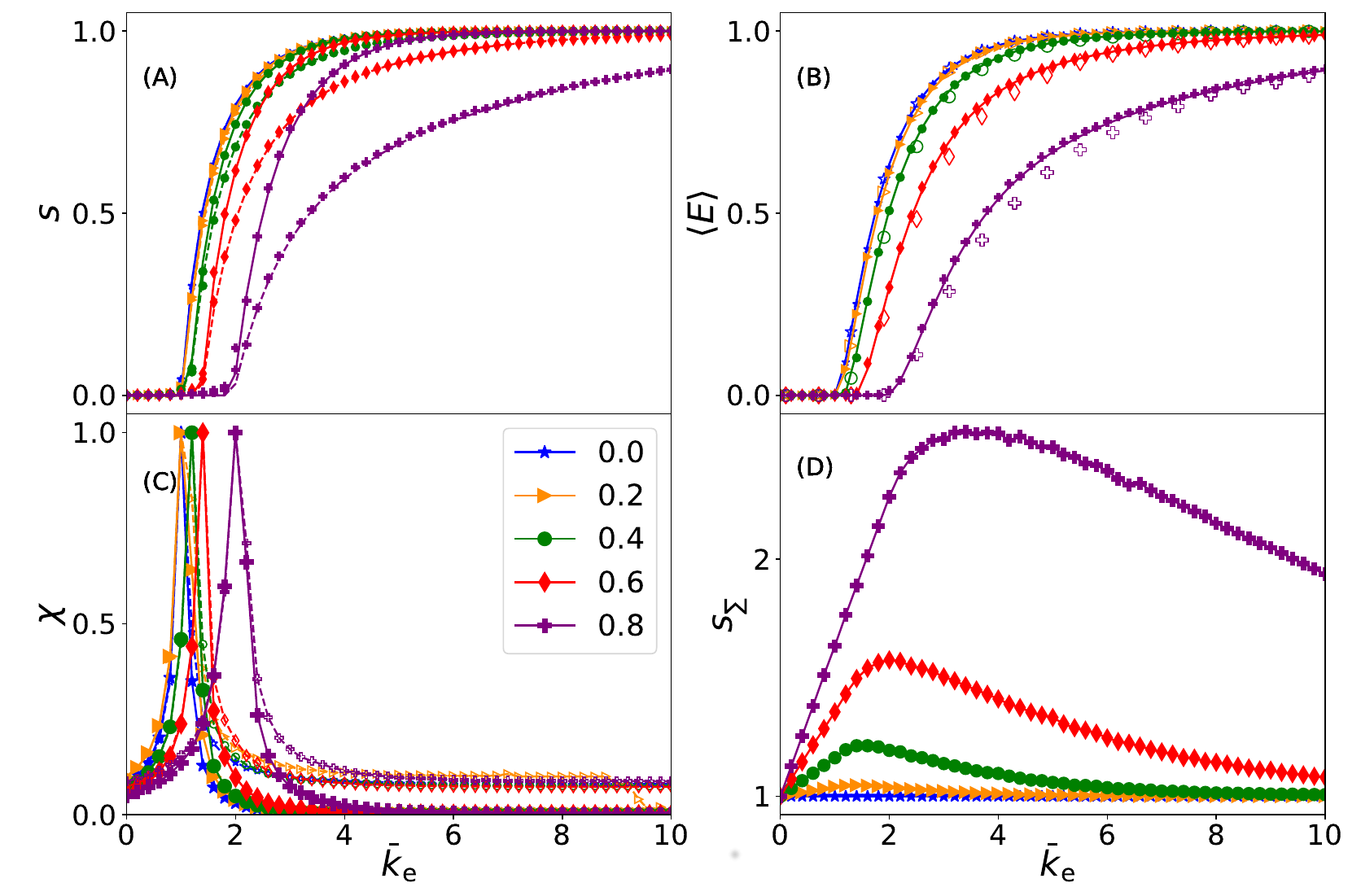}
    \caption{Disease spreading statistics in an Erd\H{o}s-R\'enyi network as a function of the effective connectivity $\bar{k}_{\rm e}$
    when there are $\pi_{\rm a}N$ perfect applications ($p_{\rm app} = 1$) that are distributed uniformly randomly. Results are normalised to the network size $N$ and shown for $\pi_{\rm a} \in [0, 0.2, 0.4, 0.6, 0.8]$ with different markers. 
    (A) The normal component size, i.e., the epidemic probability (dashed lines and markers following them) and the extended components, i.e., the epidemic size (solid lines and markers following them).  Dashed and solid lines indicate the results from theory introduced in Sec.~\ref{sec:Ali} by Eq.~\ref{eq:normal_comp_size_theory} and the markers are results computed from component sizes of simulated networks as described in Sec.~\ref{sec:Abbas}. (B) The expected epidemic size as given by Eq.~\ref{eq:expected_size} computed with theoretical results introduced in Sec.~\ref{sec:Ali} (solid lines), simulated component sizes introduced in Sec.~\ref{sec:Abbas} (filled markers), and explicit SIR simulations introduced in Sec.~\ref{sub:arashSIR} (empty markers).
    (C) Susceptibility of the normal giant component $\chi$ (dots) and the extended component $\chi'$ (solid lines) as defined in Eqs.~(\ref{Susceptibility1})-(\ref{Susceptibility2}). Since susceptibility is a divergent quantity at the epidemic threshold, as explained in Sec.~\ref{Susceptibility_formula}, it is a good proxy for finding the critical point. Notice that peaks are at the same positions for both curves, normal and extended components. (D) The fraction of sum of component sizes and network size $S_{\sum}/N$.}
    \label{fig:1}
\end{figure}

The difference between the epidemic probability (normal component size) and the epidemic size (extended component size), as given by Eqs.~(\ref{eq:normal_comp_size_theory}) and (\ref{eq:ex_comp_size_theory}), is a phenomenon specific to epidemics in the presence of app-adaptors. Breaking the equivalence of these two measures can have practical consequences, as illustrated in Fig.~\ref{fig:1}A. The difference between these two grows with the fraction of app-users $\pi_{\rm a}$. For example, when $\pi_{\rm a}=0.8$ and the epidemic probability (the normal component size) is $s_{\max} \approx 0.5$, the epidemic size (the extended component size) reaches $s_{\max} \approx 0.8$. This is also reflected in the expected epidemic sizes (see Fig.~\ref{fig:1}B and Eq. (\ref{eq:expected_size}). Despite the two component definitions differing from each other, they still display the transition at the same point and this point can be measured numerically using the susceptibilities defined in Eqs.~(\ref{Susceptibility1})-(\ref{Susceptibility2}) (see Fig.~\ref{fig:1}C). 

The extended component size is not a conserved quantity like the normal component size in the sense that the sum of component sizes $S_{\sum}$ would always sum to the number of nodes $N$. Instead, the sum of component sizes can be significantly larger than the number of nodes (see Fig.~\ref{fig:1}D) and the maximum value it can reach grows with the number of application users $\pi_{\rm a}$. The deviation from $S_{\sum}/N=1$ reaches its maximum with disease parameters higher than the threshold values, but when the disease reaches a large enough population, the fraction $S_{\sum}/N$  starts to decay, reaching $S_{\sum}/N=1$ when everybody belongs to the normal giant component.

\subsection{Quarantine failures}\label{sec:failures}
 \begin{figure}[!htb]
     \bigskip   
 \includegraphics[width=.45\linewidth]{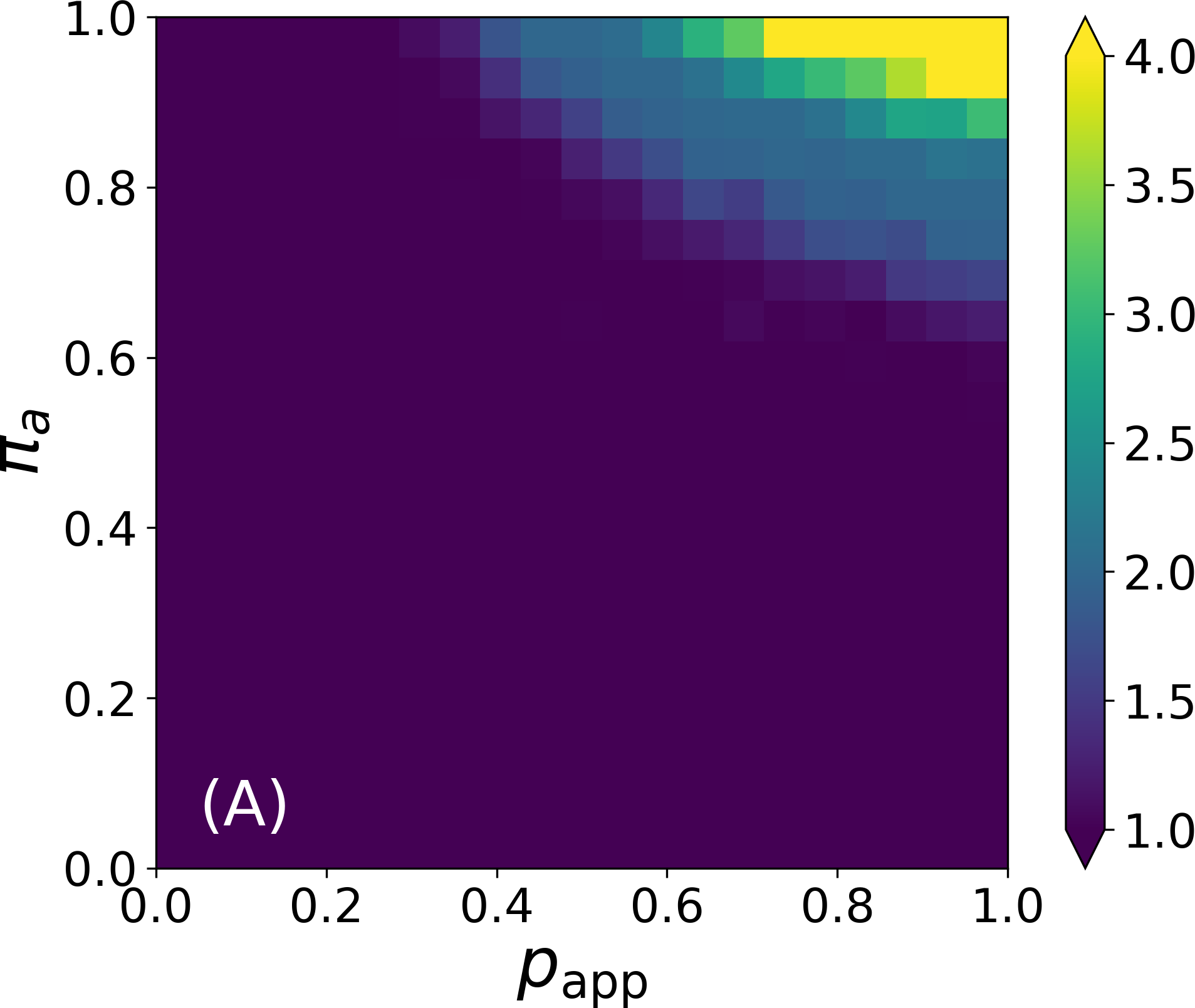}%
  \hspace{0.2cm}  
 \includegraphics[width=.45\linewidth]{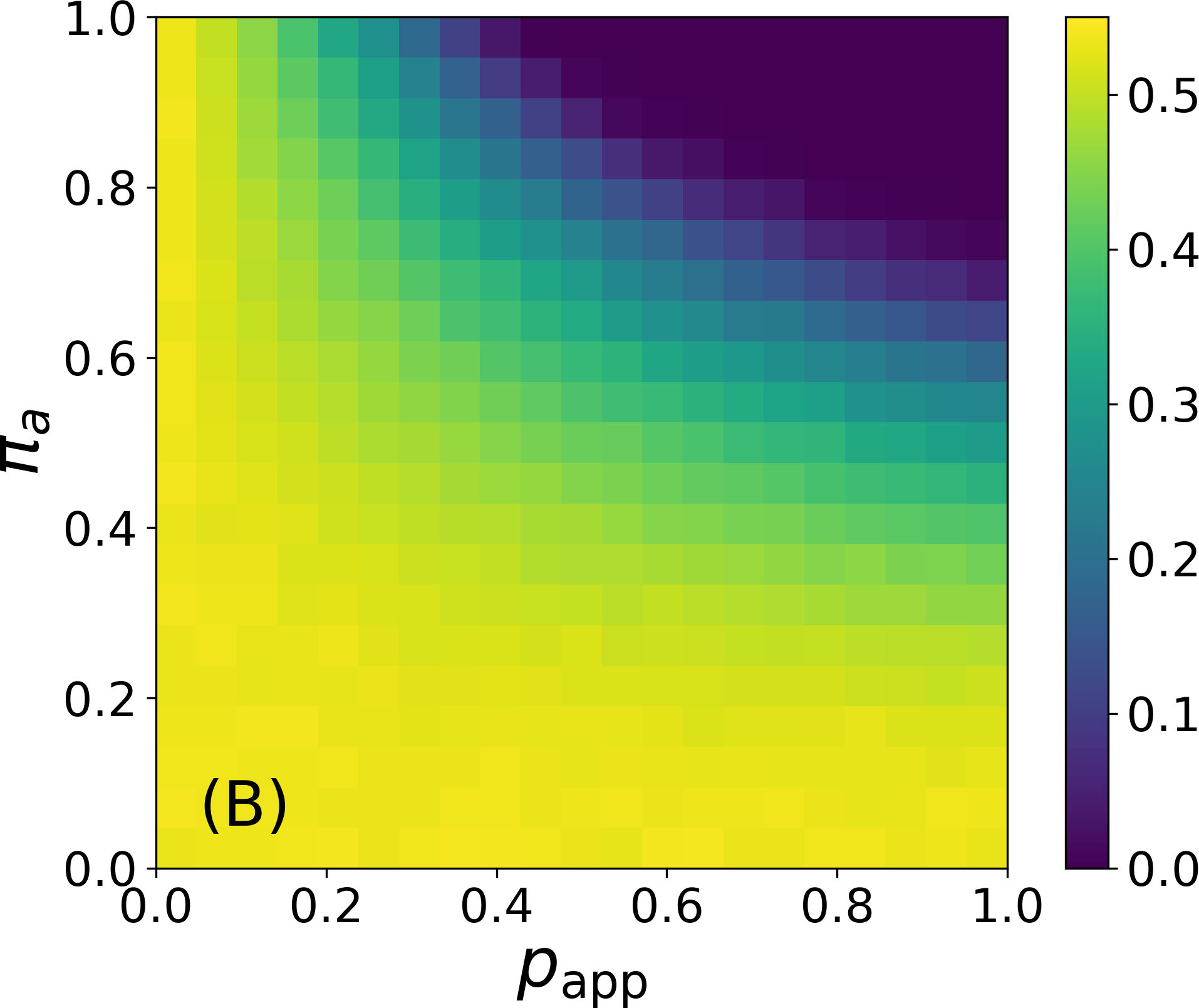}
 
  \vspace{0.5cm}
  
 \includegraphics[width=.45\linewidth]{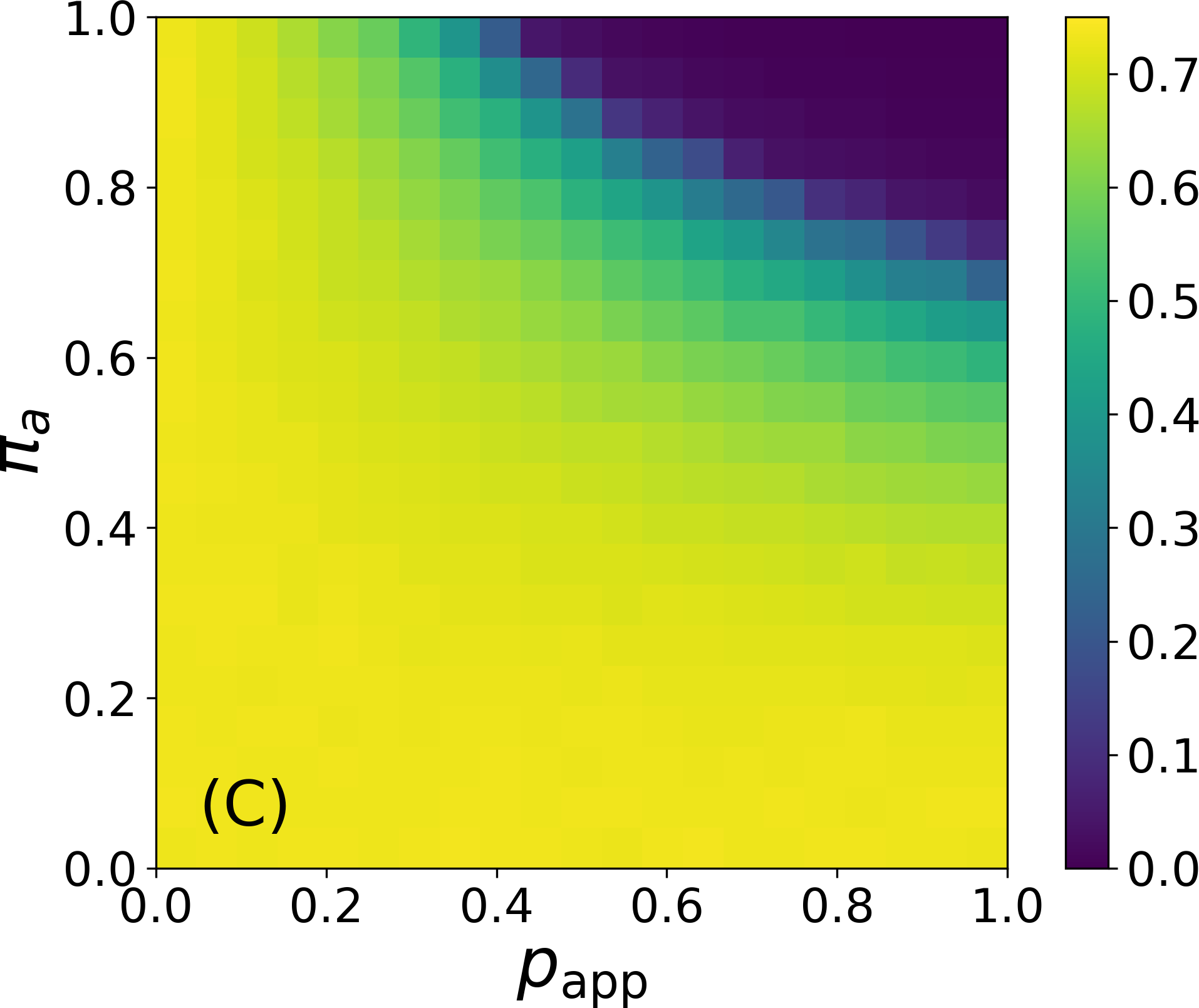}%
 \hspace{0.2cm}  
 \includegraphics[width=.45\linewidth]{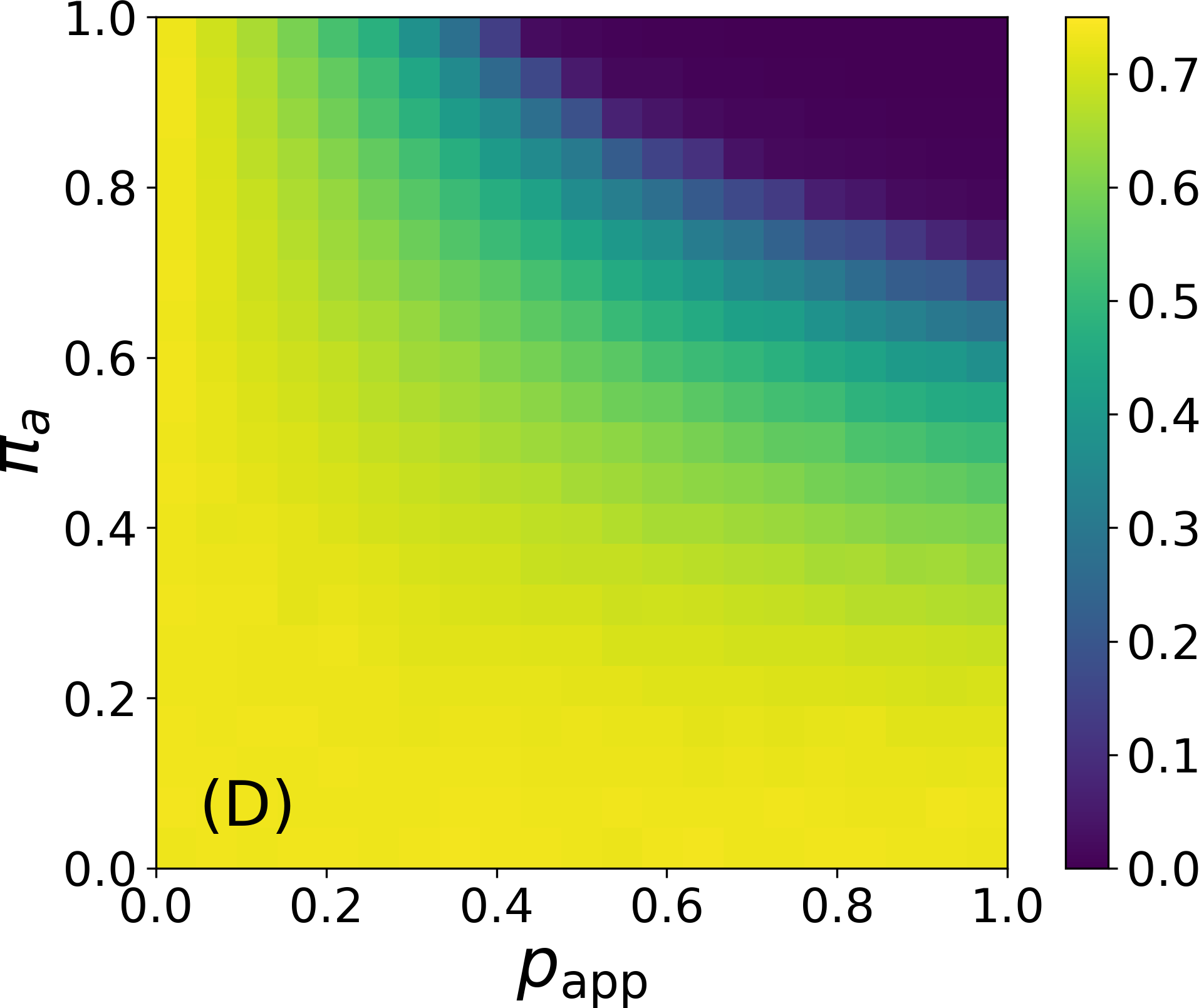}
    \caption{The effect of quarantine failures as described in Sec.~\ref{sec:failures} in homogeneous networks when app adoption is done uniformly randomly. Results are from percolation simulations. (A) The epidemic threshold as a function of quarantine probability $p_{\rm app}$ and app adoption rate $\pi_{\rm a}$. All threshold values larger than $4$ are shown with the same color. By setting the effective connectivity of the network to $\bar{k}_{\rm e} = 1.8$ (B) the expected epidemic size, (C) the extended giant component size and (D) the normal giant component size are shown as a function of $p_{\rm app}$ and $\pi_{\rm a}$. 
    Note that $\bar{k}_{\rm e} = 1.8$ is chosen as an illustrative example of a parameter region with interesting behavior in the various component sizes: it is large enough such that without any intervention, there is a wide epidemic spreading, but small enough such that the spread can be controlled without extreme measures.}
    \label{fig:2}
\end{figure}

The assumption in Section~\ref{sec:ResultsGIANT} is that i) apps work perfectly and ii) an app-user always self-isolates before having a chance to spread the infection, meaning that there are no quarantine failures, $p_{\rm app} = 1$. It is of practical significance to investigate the effects of quarantine failures \cite{wymant2021epidemiological} on the epidemic threshold and epidemic size. Fig.~\ref{fig:2} shows that in the absence of major quarantine failures, epidemic tracing and mitigation with apps can still be a valid strategy if the app adoption level in a society is high enough.
The effect of app adoption rate $\pi_{\rm a}$ is more important than the rate at which apps function, but both need to be relatively high in order for the apps to have a significant impact.

Even if we are above the epidemic threshold, the apps can be useful. Especially when the application adoption $\pi_{\rm a}$ is high, the quarantines can be very unreliable and the outbreak size (Fig.~\ref{fig:2}B-C) and epidemic probability  (Fig.~\ref{fig:2}D) both remain small. Again, overall, app adoption and quarantine reliability are essential, with the app adoption rate being more important. 

\subsection{Degree heterogeneity and high-degree app targeting}

\begin{figure}[!htb]
\label{fig:homogetreo}
 \includegraphics[width=1\linewidth]{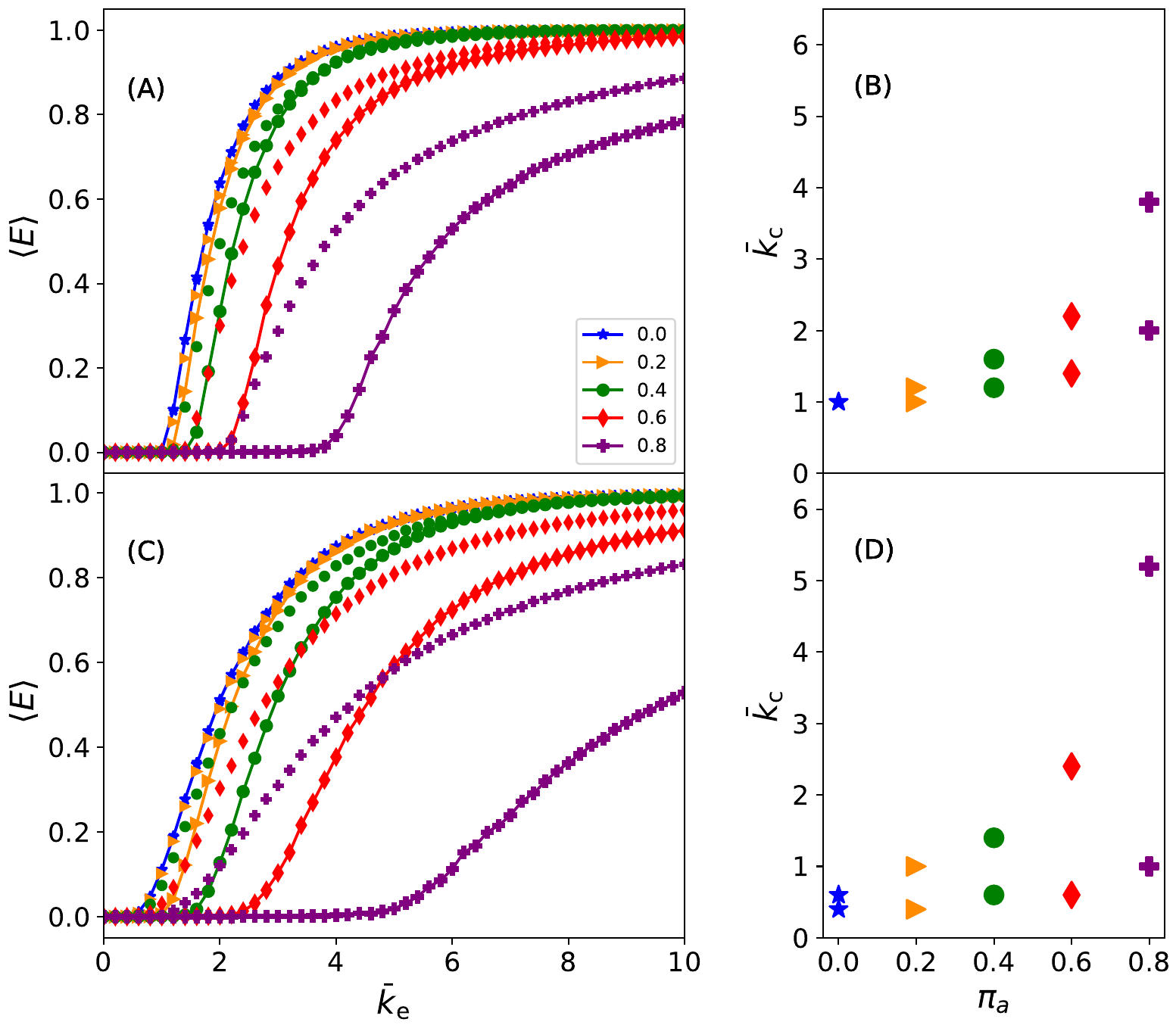}
    \caption{Expected epidemic size  $\langle E \rangle$ and epidemic threshold $\bar{k}_{\rm c}$ for two network topologies with different strategies; $\langle E \rangle$ as a function of effective connectivity $\bar{k}_{\rm e}$ for (A) homogeneous networks with Poisson degree distribution and  for (C) heterogeneous networks with a power-law degree distribution $P(k)\propto k^{-3}$. Results are shown for different values of $\pi_{\rm a}$ using different markers: $0$ (stars), $0.2$ (triangles), $0.4$ (discs), $0.6$ (diamonds), and $0.8$ (crosses). The solid lines with markers indicate the high-degree targeting strategy, while single markers indicate the random app adoption. Epidemic threshold $\bar{k}_{\rm c}$ as a function of app-adoption rate $\pi_{\rm a}$ (such that the upper markers represent the high-degree targeting strategy) for (B) homogeneous networks and for (D)  heterogeneous networks. Differences between the threshold values in the presence of homophily are explained in Fig.~\ref{fig:4}B,D.}
    \label{fig:3}
\end{figure}

Real networks are degree-heterogeneous and this heterogeneity has a strong effect on the final outbreak size and the epidemic threshold. Fig.~\ref{fig:3} shows the expected epidemic sizes with two different strategies in app adoption, random  and high-degree targeting, for different fractions of app-users $\pi_{\rm a}$ in the network. In homogeneous networks, Fig.~\ref{fig:3}A, contact tracing decreases the expected epidemic size and pushes the epidemic threshold forward. These effects can be further amplified by shifting to the high-degree targeting in app adoption. With $80\%$ of app-users, the epidemic threshold can move from $\bar{k}_{\rm e} = 1$ to $\bar{k}_{\rm e} = 4$, which means at that point expected epidemic size is zero, while without contact tracing it would be almost 1. Note that in homogeneous networks, the effective average degree of the contact network $\bar{k}_{\rm e}$, has good correspondence to the reproduction number of the infection.

In networks with degree-heterogeneity, the epidemic threshold vanishes in normal SIR processes. This effect holds in contact-traced epidemics if we distribute the apps uniformly randomly. However, from Fig.~\ref{fig:3}B it is clear that contact tracing can significantly reduce the expected epidemic size even when the apps are randomly distributed and the epidemic threshold remains unchanged. With the high-degree targeting strategy, it is possible to move the epidemic threshold. Comparing the expected epidemic size at different values of $\bar{k}_{\rm e} < 3$ shows that in real-world situations, app adoption of superspreaders is of significant importance. Since hubs become the app-users, this strategy has drastic effects on the size and threshold of the epidemic, such that the threshold gets pushed from somewhere near zero to a value $\bar{k}_{\rm e} > 5$ with the app adoption rate $\pi_{\rm a}=0.8$. Therefore, the reproduction number can be much more controlled in the high-degree targeting strategy.

\subsection{The effect of homophily and heterophily}
\label{sec:homophily}

 \begin{figure}[!htb]
    \bigskip
    \includegraphics[width=.45\linewidth]{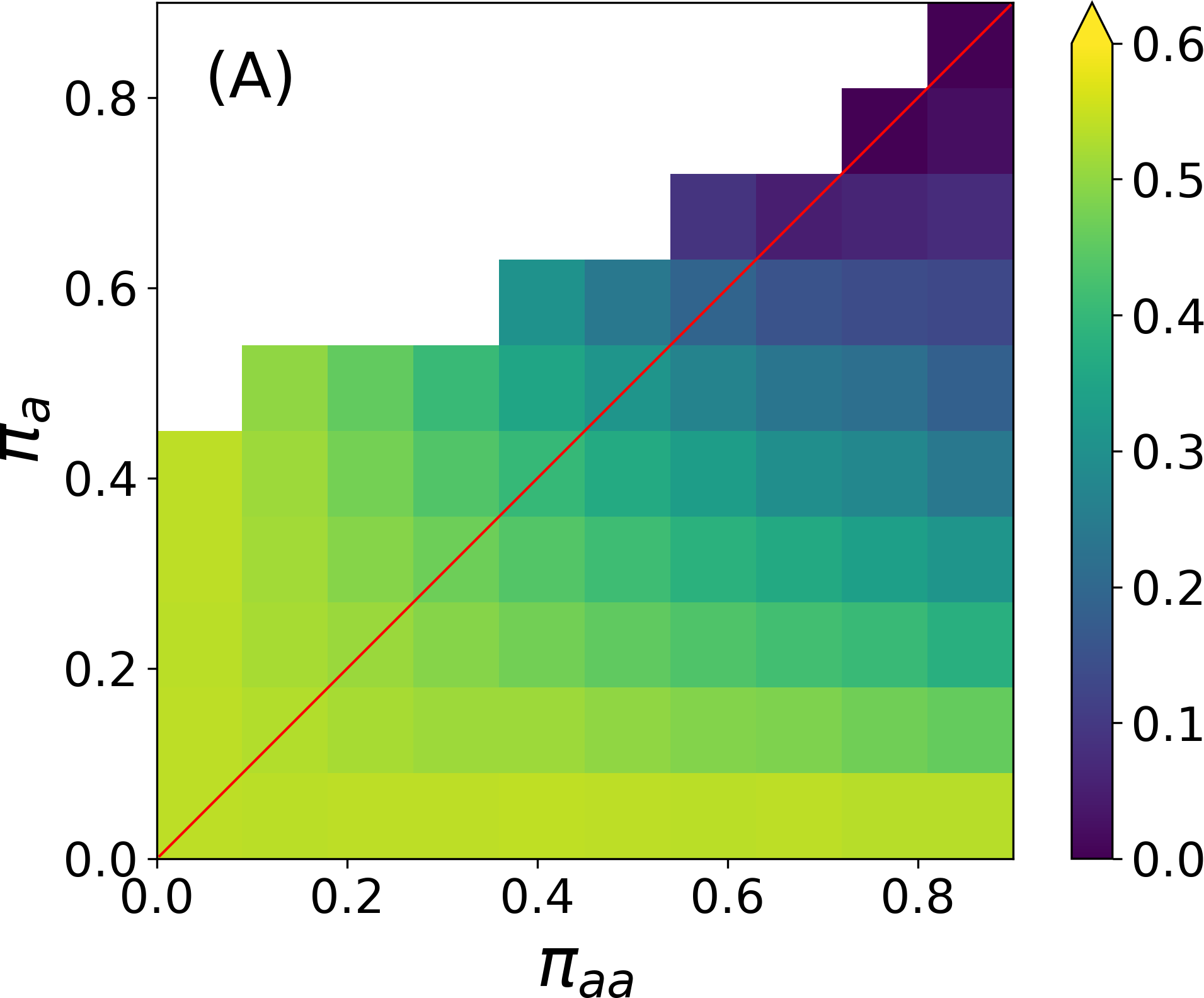}%
    \hspace{0.2cm}
    \includegraphics[width=.45\linewidth]{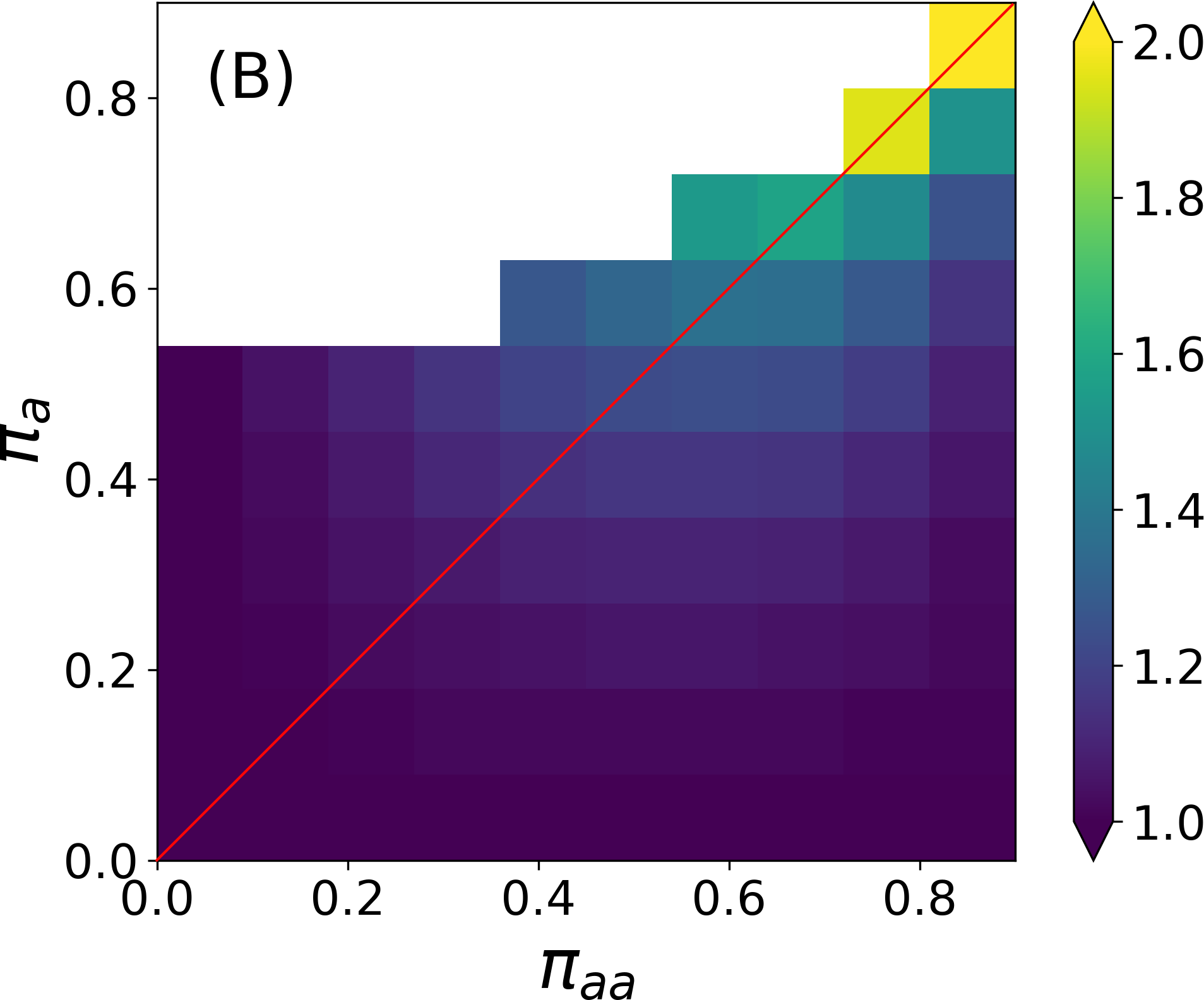}
    
    \vspace{0.5cm}
    \includegraphics[width=.45\linewidth]{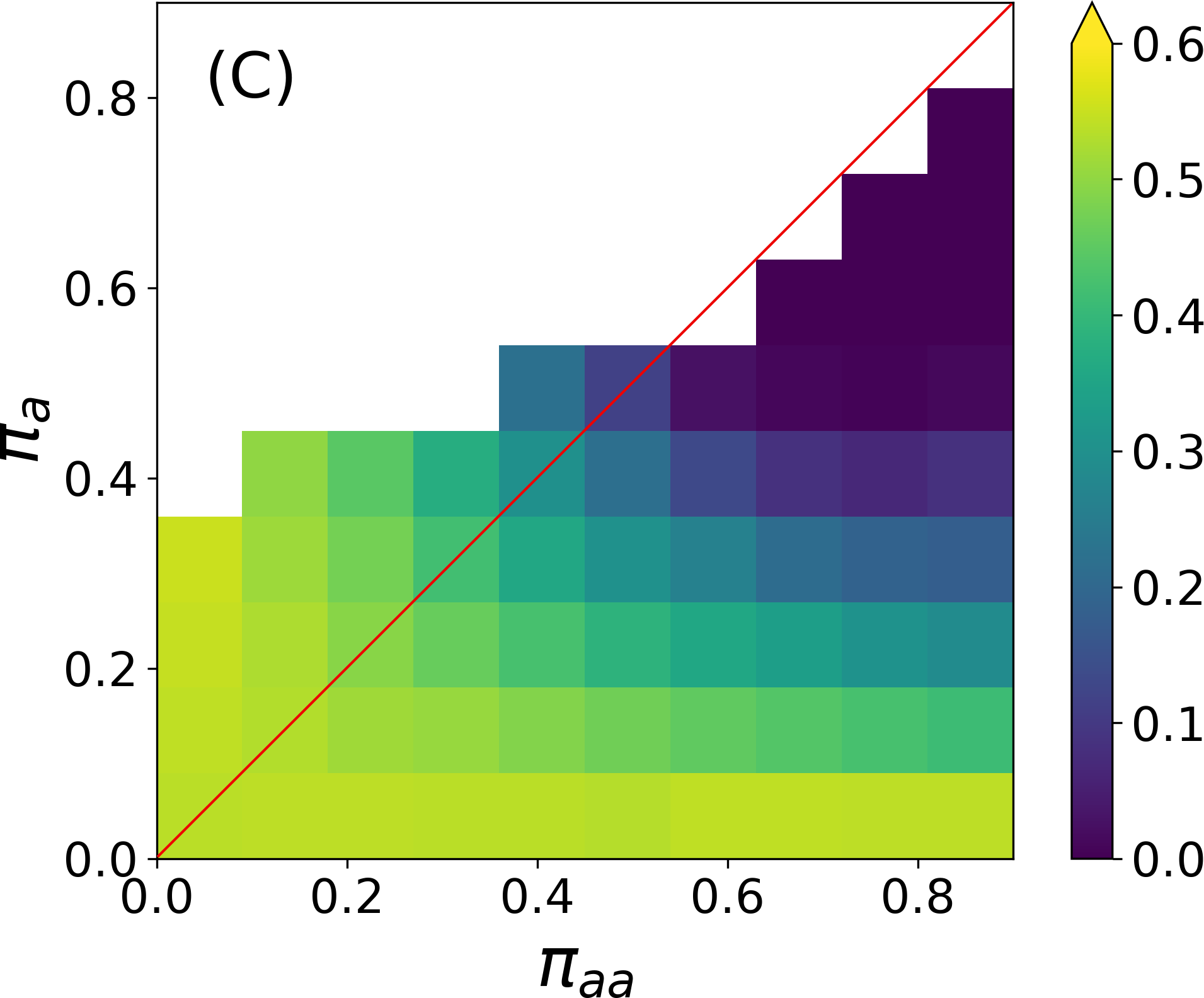}%
     \hspace{0.2cm}
     \includegraphics[width=.45\linewidth]{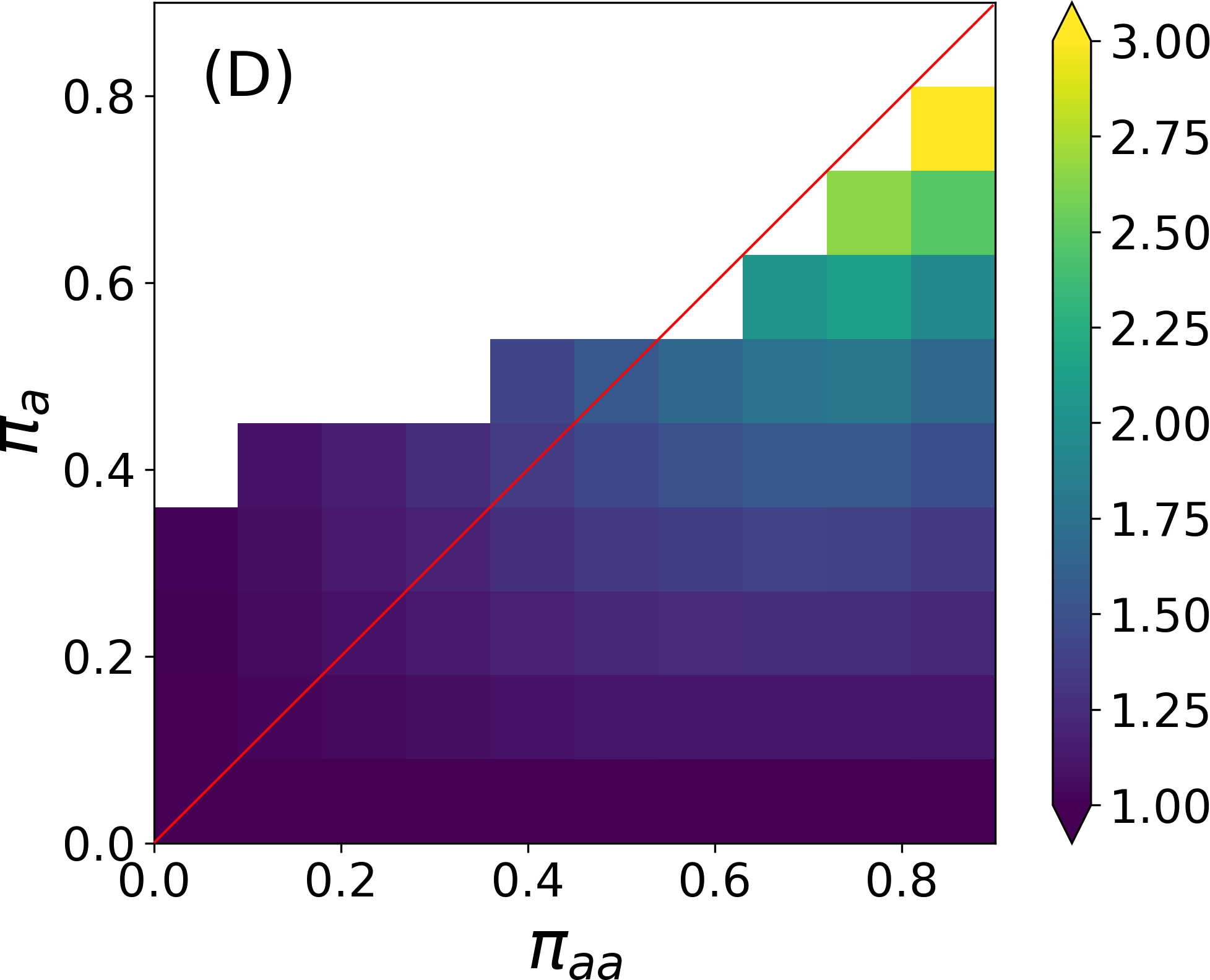}
    \caption{The effect of homophily/heterophily in app adoption in homogeneous networks  as described in Sec.~\ref{sec:homophily}. Homophily (heterophily) region is below (above) the diagonal $\pi_{\rm a}=\pi_{\rm aa}$. Expected epidemic size at $\bar{k}_{\rm e} = 1.8$ for (A) random app adoption and for (C) high-degree targeting strategy. The epidemic threshold for (B) random app adoption and for (D) the high-degree targeting strategy. Thresholds are from theoretical results given by Eq.~\ref{kc_homo} and expected epidemic sizes are from percolation simulations.
    The empty white region is the spectrum that having such a homo/heterophilic population is impossible.}
    \label{fig:4}
\end{figure}

In previous sections, there was an assumption that app-users are distributed with random mixing patterns; the fact that one of the connections of a node is an app-user has no effect on the probability of that node being an app adopter. Next, we explore how homophily/heterophily affects epidemics based on app usage using the modular network model (MN). A Swiss experiment has reported that while a small fraction of $\pi_{\rm a} = 0.2$ of people have used the app, the inside connections between them was high enough such that $\pi_{\rm aa} = 0.7$ \cite{salathe2020early}.

\begin{figure}[!htb]
\includegraphics[width=.71\linewidth]{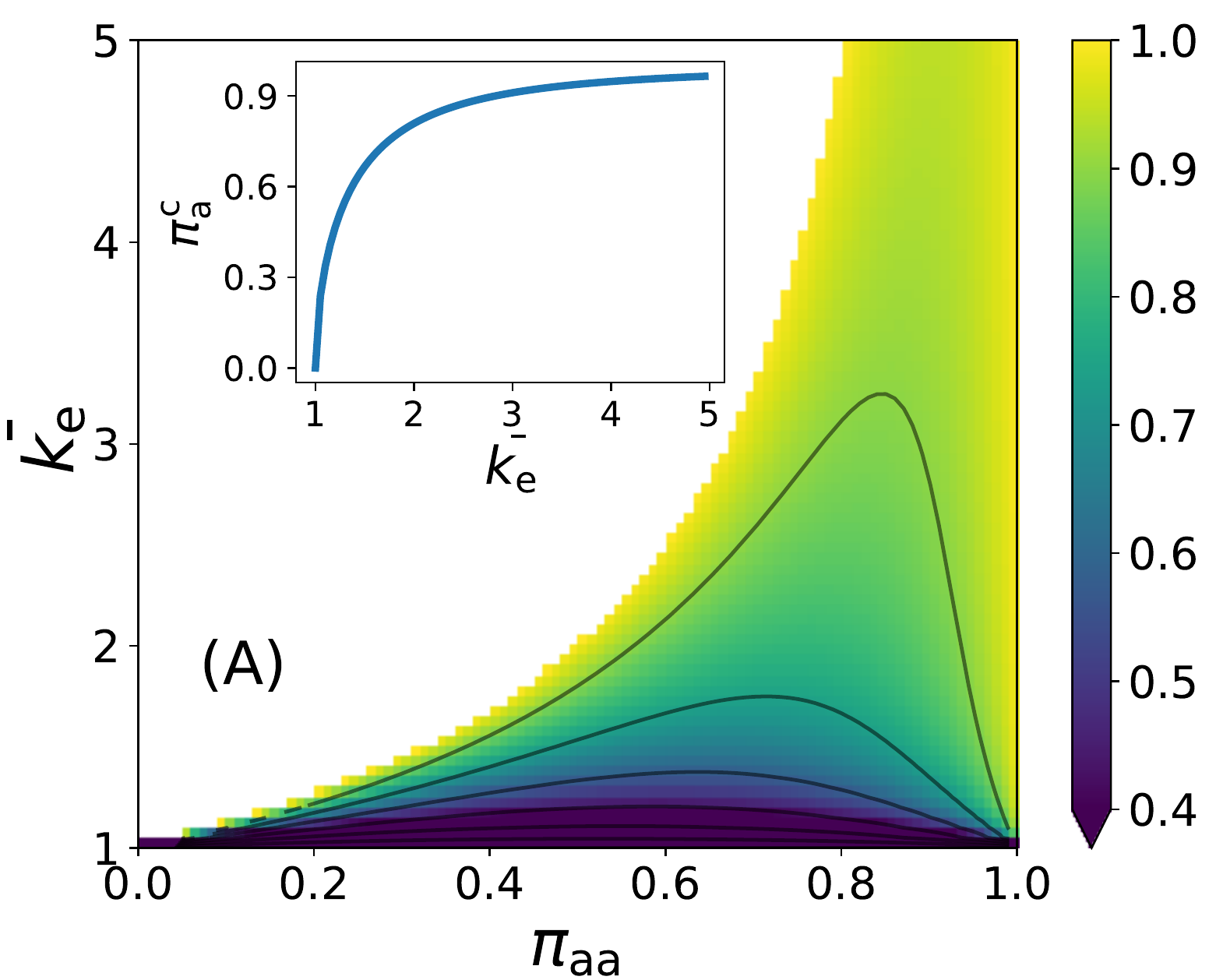}
 \vspace{0.4cm}
 \includegraphics[width=.75\linewidth]{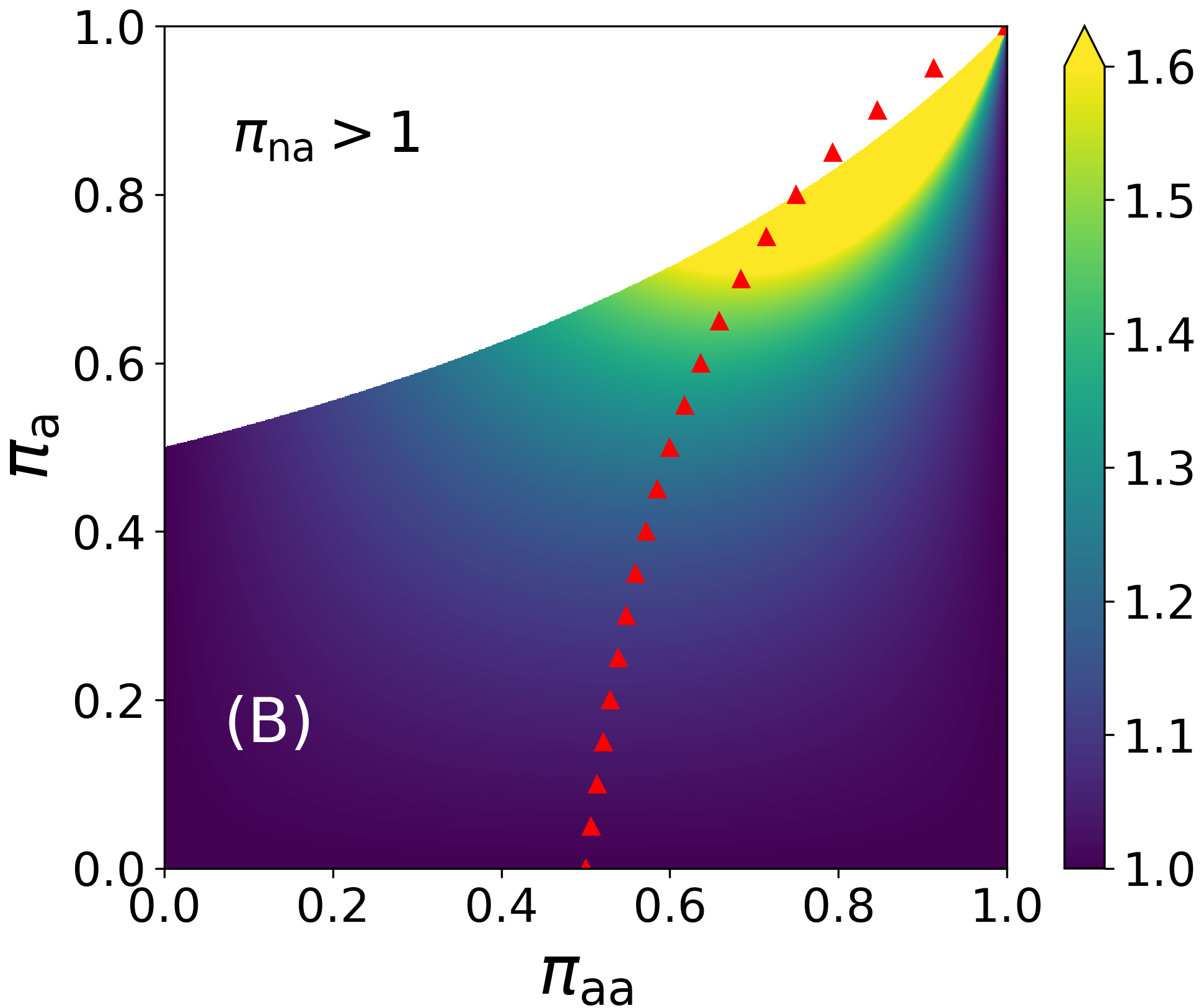}
\caption{Existence of optimum value for homophily based on branching process approximation as described in Sec.~\ref{sec:branching}. (A) The critical value of app-users $\pi_{\rm a}^{\rm c}$ that are needed for reducing the reproductive number as a function of effective connectivity and homophily probability $\pi_{\rm aa}$. The value of $\pi_{\rm a}^{\rm c}$ remains the same within each black curve. The inset is the graph of $\pi_{\rm a}^{\rm c}$ as a function of $\bar{k_{\rm e}}$ in the absence of homophily  $\pi_{\rm aa} = \pi_{\rm a}$ given by Eq.~\ref{equ:pi_c}. (B) The epidemic threshold $\bar{k_{\rm e}}$ as a function of $\pi_{\rm aa}$ and $\pi_{\rm a}$.
The red symbols show the $\pi_{\rm aa}^{\rm opt}$ for each $\pi_{\rm a}$ which is given by to Eq.~\ref{opt_po}. The pattern here is consistent with another approximation shown in Fig.~\ref{fig:4}B, while epidemic threshold values are slightly different due to different levels of approximations. Note that here we display the epidemic threshold for all values of $\pi_{\rm aa}$ and $\pi_{\rm a}$ such that $0\leq\pi_{\rm na}\leq1$ so the networks with some of these parameters can be created in practice \cite{Gleeson08}.}
   \label{fig:4_branching}
\end{figure}

Fig.~\ref{fig:4} illustrates that increasing heterophily leads to a lower epidemic threshold and larger epidemic size for a fixed $\bar{k}_{\rm e}$. Increasing homophily from random mixing is initially preferable, but the optimum lies between random mixing and full homophily. 
For the expected epidemic size, strong heterophily is especially detrimental (see Fig.~\ref{fig:4}A for the homogeneous network and with random app adoption and in Fig.~\ref{fig:4}C for high-degree targeting strategy). 
The optimum value for heterophily/homophily is evident for the epidemic thresholds in  Fig.~\ref{fig:4}B and Fig.~\ref{fig:4}D, respectively, for the random and high-degree targeting strategies. 
Fig.~\ref{fig:4_branching}B gives a more clear picture of existence of an optimum value for the epidemic threshold in the case of homophily. According to Eq.~\ref{kc_homo}, for each fraction of app-users $\pi_{\rm a}$ in the network, the epidemic threshold $\bar{k}_{\rm c}(\pi_{\rm a},\pi_{\rm aa})$ can be maximised by controlling the homophily in app adoption $\pi_{\rm aa}$. The pattern in the Fig.~\ref{fig:4_branching}B is very similar to the convex pattern in Fig.~\ref{fig:4}B, even though they are calculated using different approximations and approaches (see Sec.~\ref{sec:Ali} and Sec.~\ref{sec:branching}).

Another view on the effect of homophily and heterophily is given by finding the critical fraction app-users $\pi_{\rm a}^{\rm c}$ that is needed to go beyond the epidemic threshold as a function of  $(\pi_{\rm aa}$ and $\bar{k}_{\rm e})$. Fig.~\ref{fig:4_branching}A  depicts this relationship based on Eq.~\ref{abbas_pc} and shows that  $\pi_{\rm a}^{\rm c}$ is not monotonic function of $\pi_{\rm aa}$ but there is an optimal value of $\pi_{\rm aa}$ giving the lowest fraction apps that are needed to stop the epidemic.
Note that in a network without homophily or heterophily $\pi_{\rm a}^{\rm c}$ increases monotonically as the function of the effective connectivity $\bar{k}_{\rm e}$ (see the inset of Fig.~\ref{fig:4_branching}a).

\section{Discussions}

In this article, we have developed two flexible analytic approximations to SIR epidemics in the presence of contact tracing apps. First, we use a branching process to derive explicit analytical solutions for the epidemic thresholds. Second,
we expand the framework of using self-consistent equations to analyze digital contact tracing \cite{Barrat2020.07.24.20159947}, which is an alternative to other approaches \cite{bianconi2021message}. Contrary to the conventional SIR spreading, a full picture of the late-state epidemics in the presence of digital contact tracing is not given by a single observable (the component size), but one also needs two variables (normal and extended component sizes). These correspond to the probability of the epidemic and the epidemic size, which are equivalent in the SIR process. Here we see that the two quantities can be significantly different if the number of application users is high.

Our numerical results illustrate that the effects of digital contact tracing can be very sensitive to the network structure, how applications are distributed among the population, and how well the tracing works. Realistic estimates of the effects of digital contact tracing can only be achieved if one can choose correct parameter ranges in a high-dimensional parameter space. In this study, we had 6 of such parameters: the shape of the degree distribution, average degree, amount of heterophily/homophily, application prevalence, quarantine probability and targeting strategy. While we were able to establish and confirm basic laws governing individual parameters and some combinations of parameters, exploring such a parameter space fully for possible compound effects is out of the reach in simulations. However, these effects can be largely revealed by inspecting the analytic equations we derived.

There are several open questions for which this study and other studies only hint at the results. There are types of network structures we ignore here. For example, the heterophily and homophily could be constructed in the network in slightly different ways. 
For example, a case study using a realistic agent-based model \cite{lopez2020anatomy} has recently considered, among many other modeling choices aimed at precise calibration on the French population, the contributions of individuals of different ages. One could also develop a more realistic version of our stylized model to systematically analyze the effects of homophily caused by an age-based contact structure and different scenarios of app adoption within that structure. The age-based approach would also allow one to estimate the benefits of applications relative to the risk groups in this model.

Overall the problem of digital contact tracing offers not only a practical problem to solve but also an interesting theoretical puzzle because it introduces memory to the epidemic process. This memory is limited to one step within the tracing model we use here, but one could also use multi-step tracing, where also the second neighbors of infected nodes are quarantined in the case that the first neighbors have already passed on the infection. Further, here we ignore effects such as quarantines that do not directly stop the infection from one application user to another from spreading further. However, in the case of a strong group structure in the network, there could be situations where a non-application user A infects application user B, who alerts another application user C, who actually gets infected by A and stops the spreading because of the quarantine. 
Analyzing such more complicated phenomena can provide challenges for network scientists for years to come.

\section*{Acknowledgement}
The simulations presented above were performed using computer resources within the Aalto University School of Science ``Science-IT'' project.
AF acknowledges funding by Science Foundation Ireland Grant No. 16/IA/4470, No. 16/RC/3918, No. 12/RC/2289P2 and No. 18/CRT/6049.

\bibliography{citations}

\section*{Appendix}
\begin{figure}[!htb]
 \includegraphics[width=.9\linewidth]{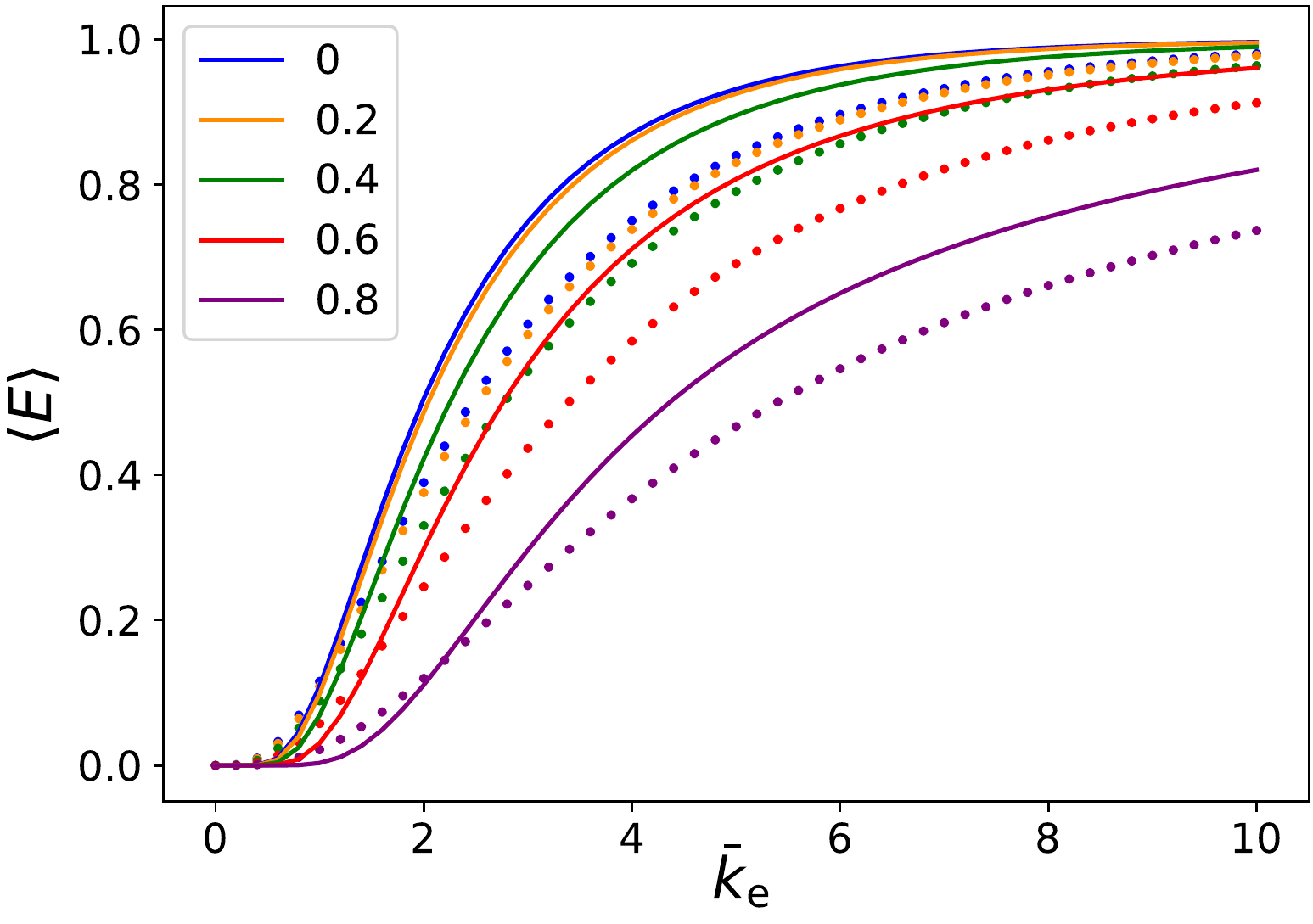}
    \caption{The expected epidemic size computed with theoretical results introduced in Sec.~\ref{sec:Ali} for heterogeneous networks with degree distribution $P(k)\propto k^{-3}$ (solid lines) compared with ones with $P(k)\propto k^{-2.5}$ (dotted lines) as a function of the effective connectivity $\bar{k}_{\rm e}$ when apps are distributed uniformly randomly. Results are normalised to the network size $N$ and shown for $\pi_{\rm a} \in [0, 0.2, 0.4, 0.6, 0.8]$ with different colors. Note that by lowering the exponent, epidemic thresholds get closer to zero and the expected epidemic sizes decrease since there more low-degree nodes in the network. Therefore, by lowering the exponent, while we can add more degree heterogeneity in the network, the physics of the phenomena does not change.}
    \label{fig:1apx}
\end{figure}

The heterogeneity in the number of contacts could also be modeled with other distributions, for example, the negative binomial distribution. This would have the advantage of having a non-divergent second moment supported by empirical evidence. However, we aimed to illustrate the effects of degree heterogeneity and not perform a systematic analysis. We already have many different random network models and combinations of parameters related to the app distribution, how well it works, and disease parameters. The equations we give make it possible for one to do such analysis if needed. Therefore we limited our main discussion to the differences observed in a power-law network with exponent $-3$ compared to the results for homogeneous networks. However, to satisfy the curiosity of the reader interested in extreme heterogeneity, we have now added Fig.~\ref{fig:1apx} showing the expected epidemic size for exponent $-2.5$ as suggested by the referee.

\begin{figure}[H]
 \includegraphics[width=.5\linewidth]{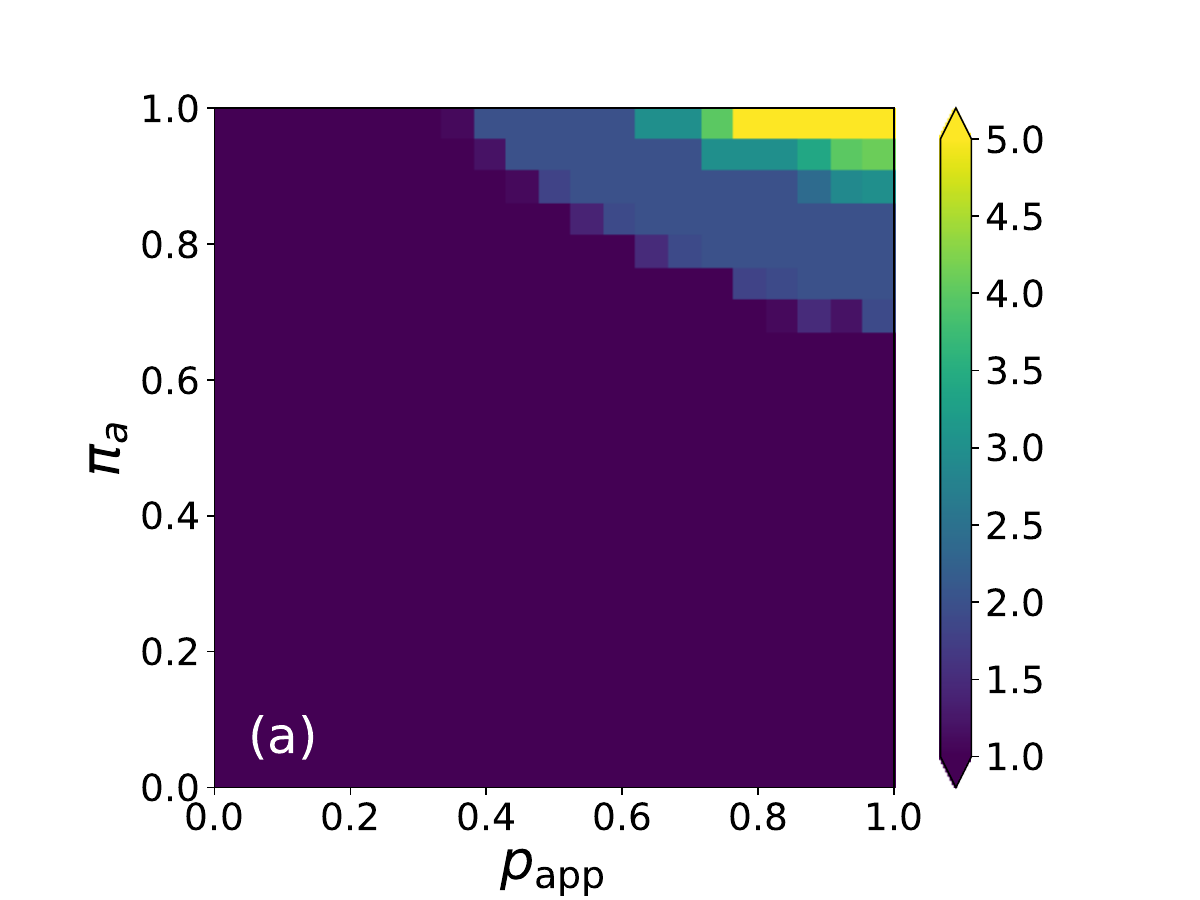}%
 \includegraphics[width=.5\linewidth]{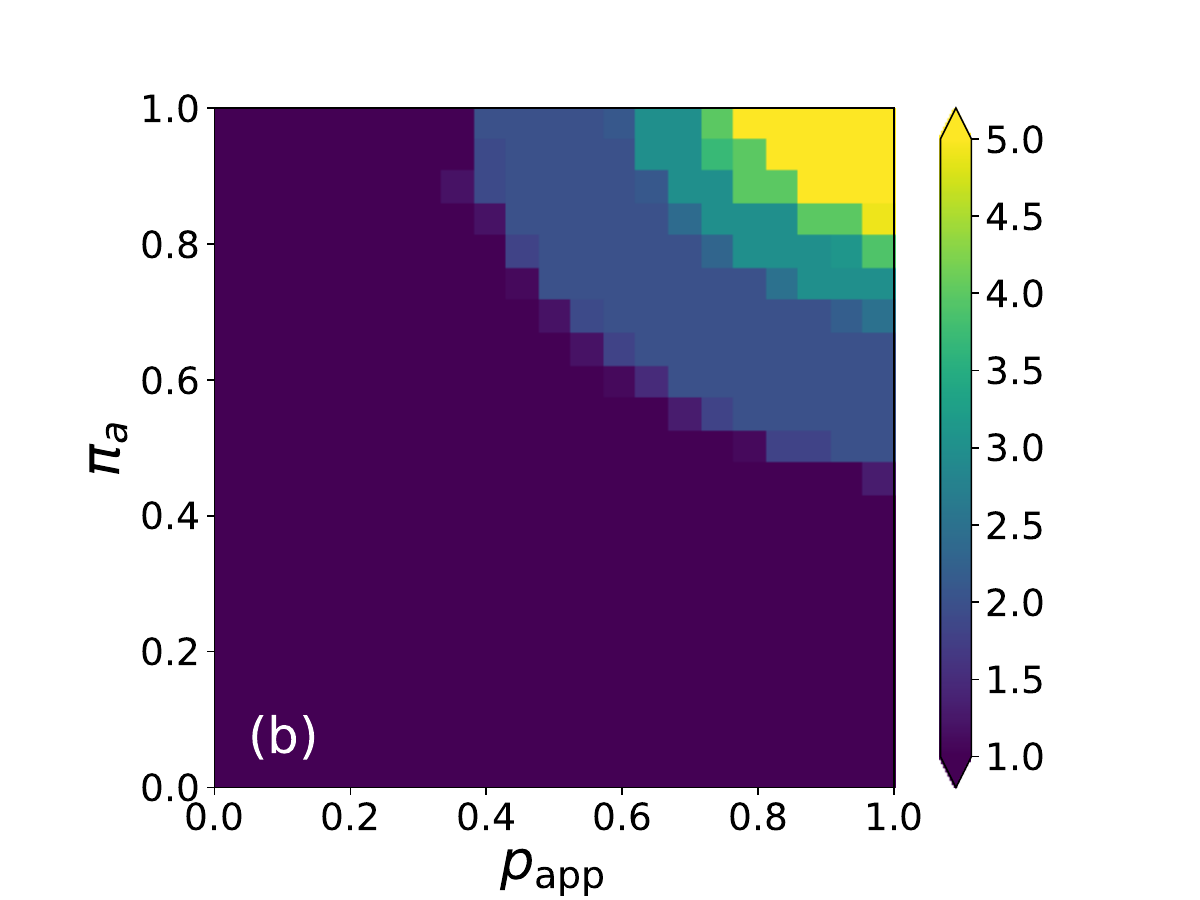}
  \vspace{0.5cm}
   \includegraphics[width=.5\linewidth]{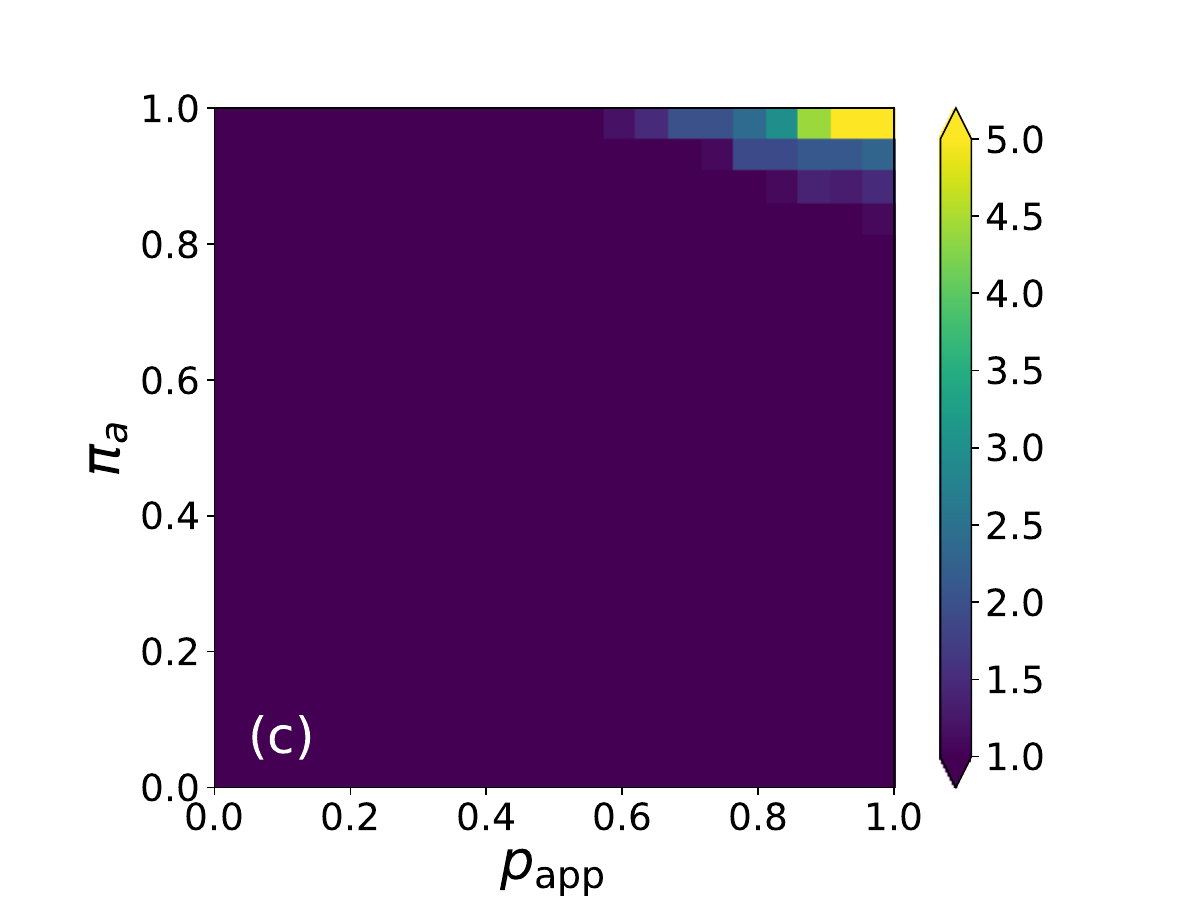}%
 \includegraphics[width=.5\linewidth]{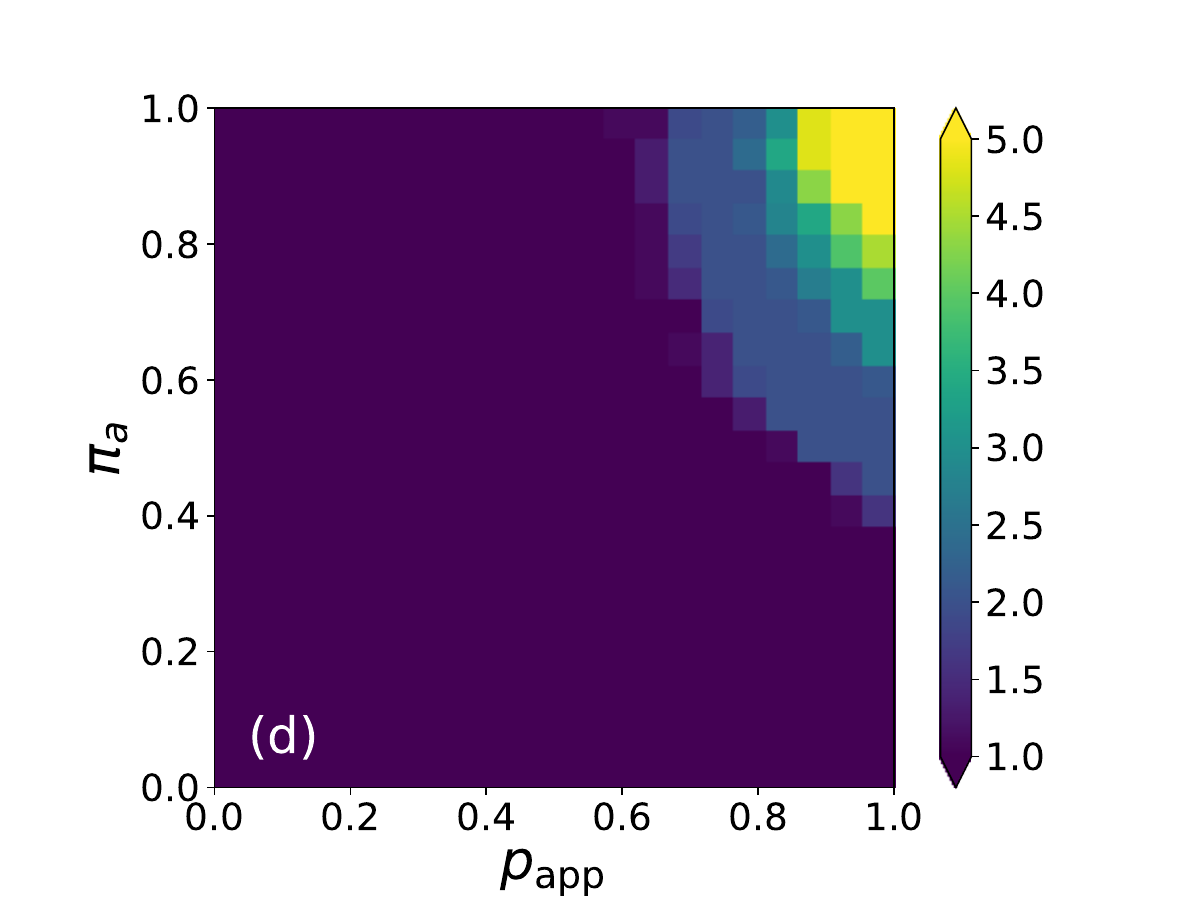}
    \caption{The epidemic threshold as a function of quarantine probability $p_{\rm app}$ and app adoption rate $\pi_{\rm a}$. The effect of quarantine failures in homogeneous networks with (a) random app adoption (b) and high-degree targeting strategy. Also, for heterogeneous networks with a power-law degree distribution with (c) random app adoption (d) and high-degree targeting strategy. 
    All threshold values larger than $5$ are shown with the same color.}
    \label{fig:2A-appendix}
\end{figure}

\begin{figure}[H]
 \includegraphics[width=.5\linewidth]{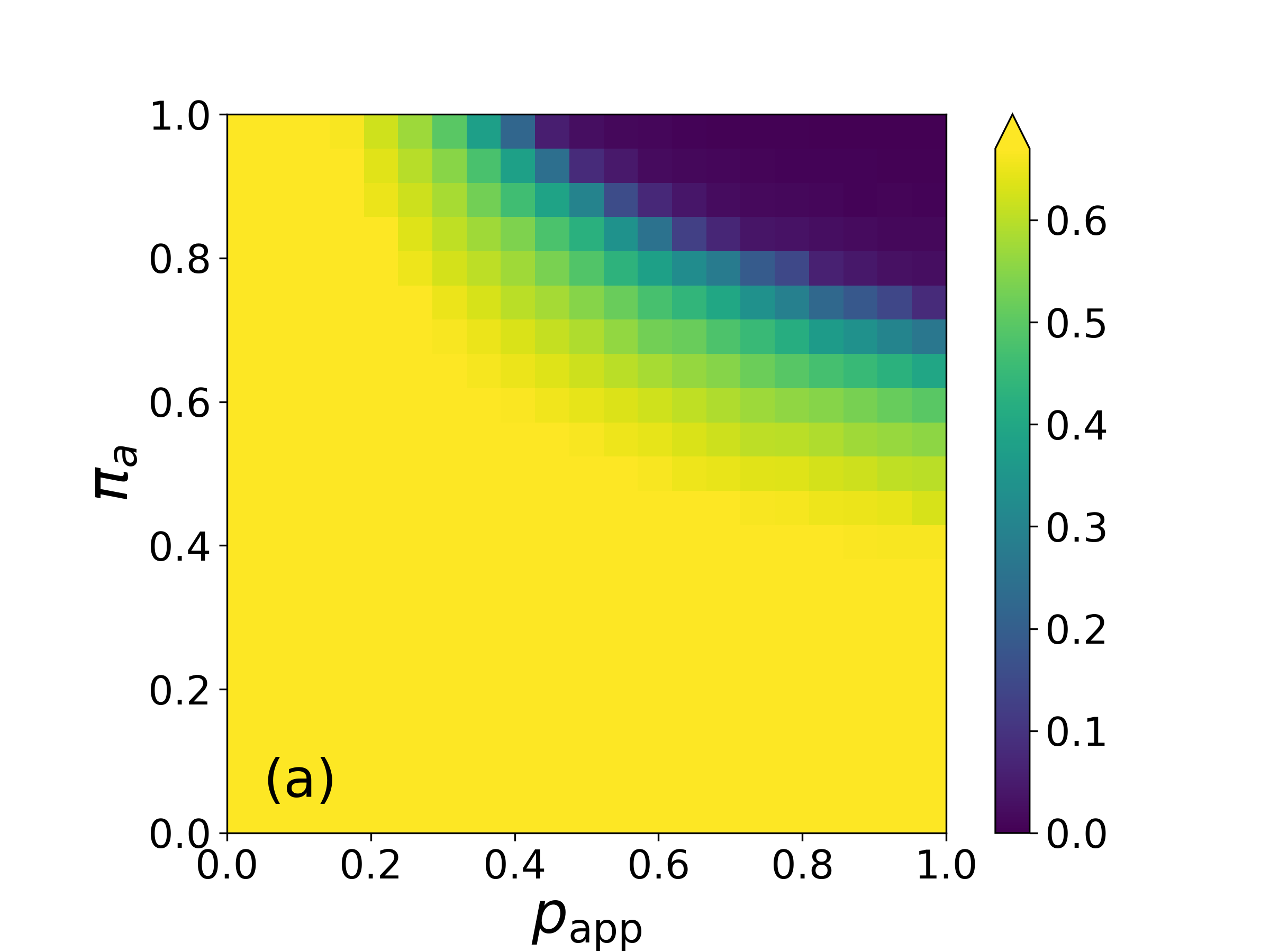}%
 \includegraphics[width=.5\linewidth]{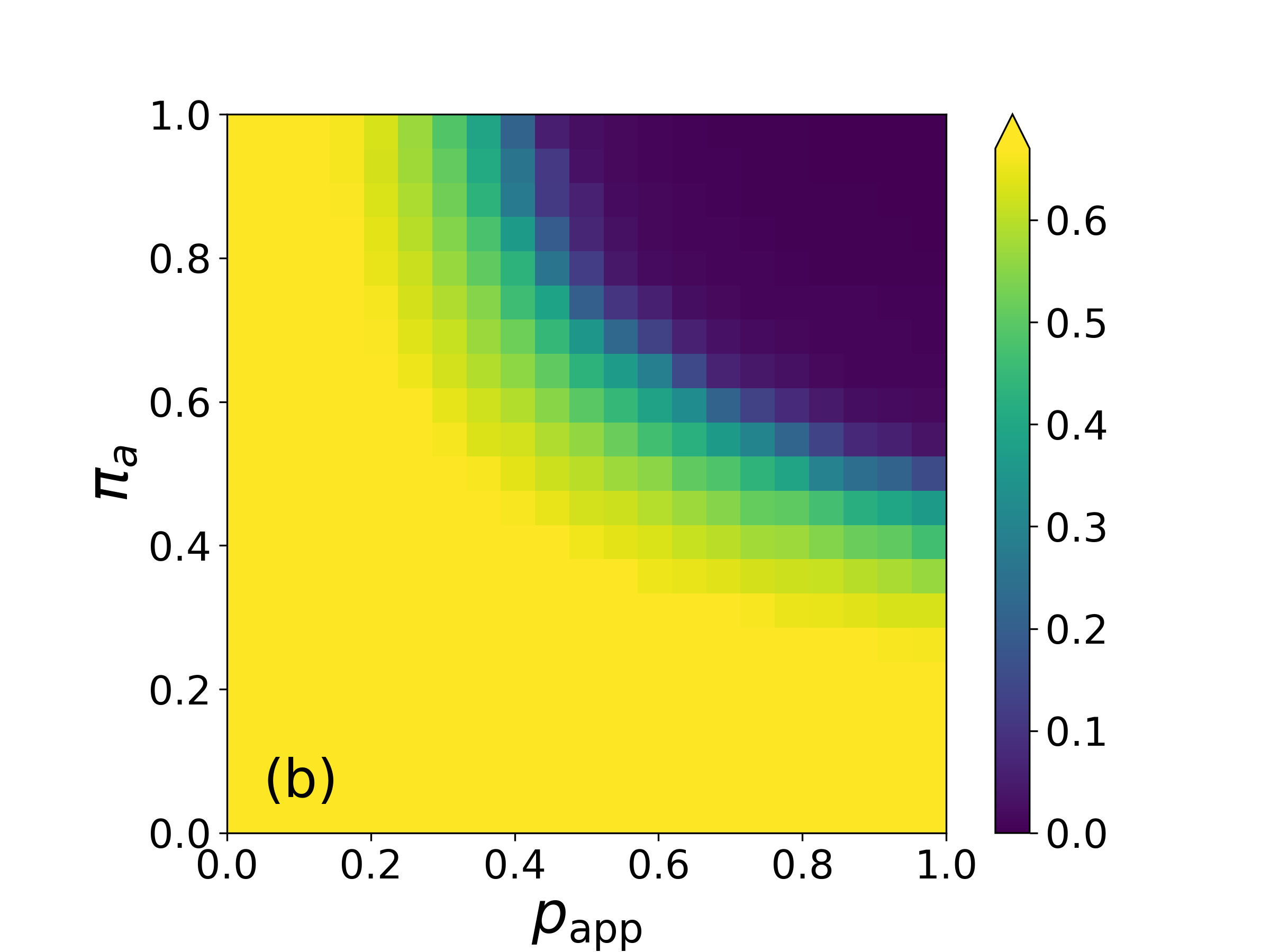}
  \vspace{0.5cm}
   \includegraphics[width=.5\linewidth]{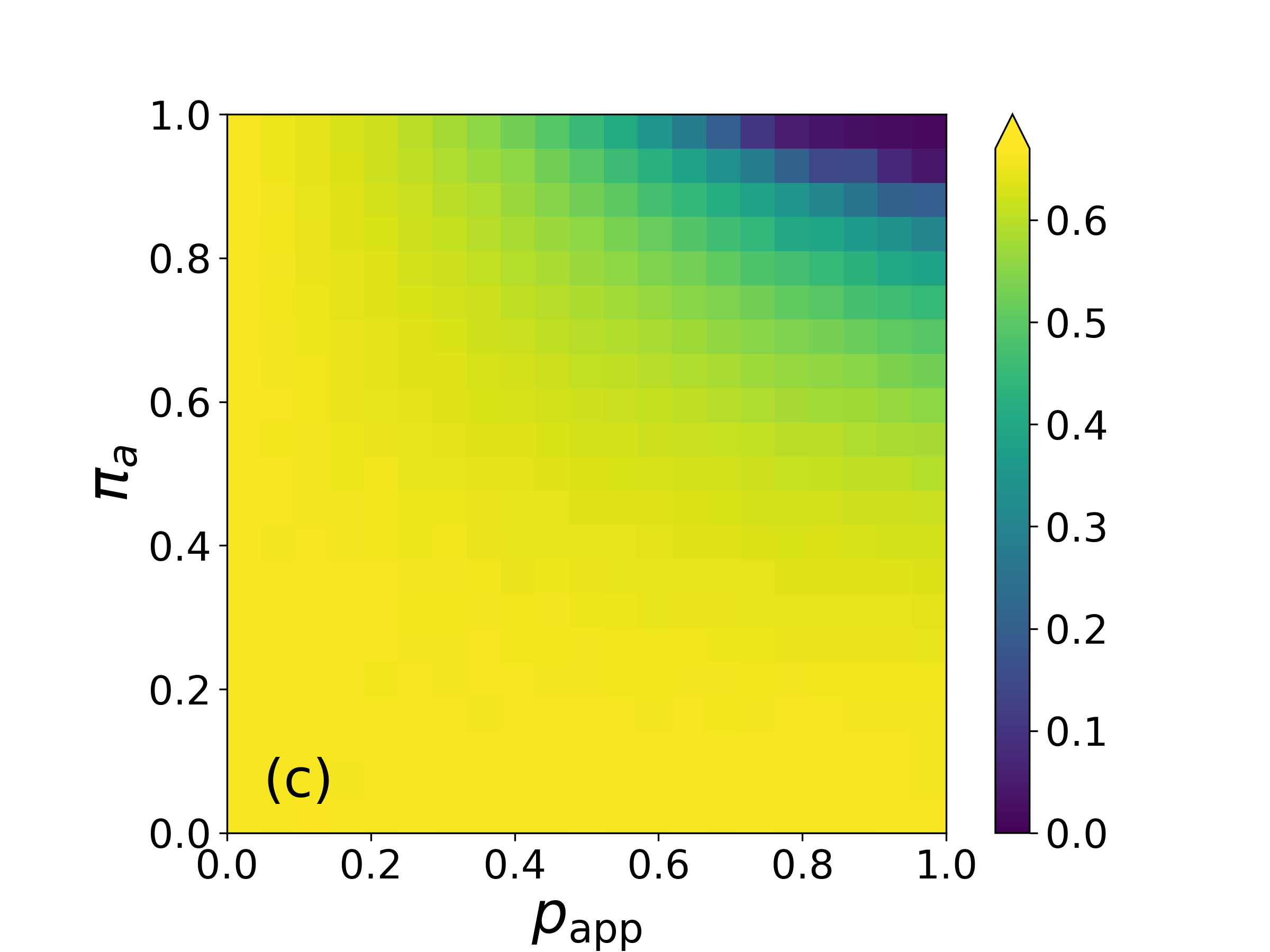}%
 \includegraphics[width=.5\linewidth]{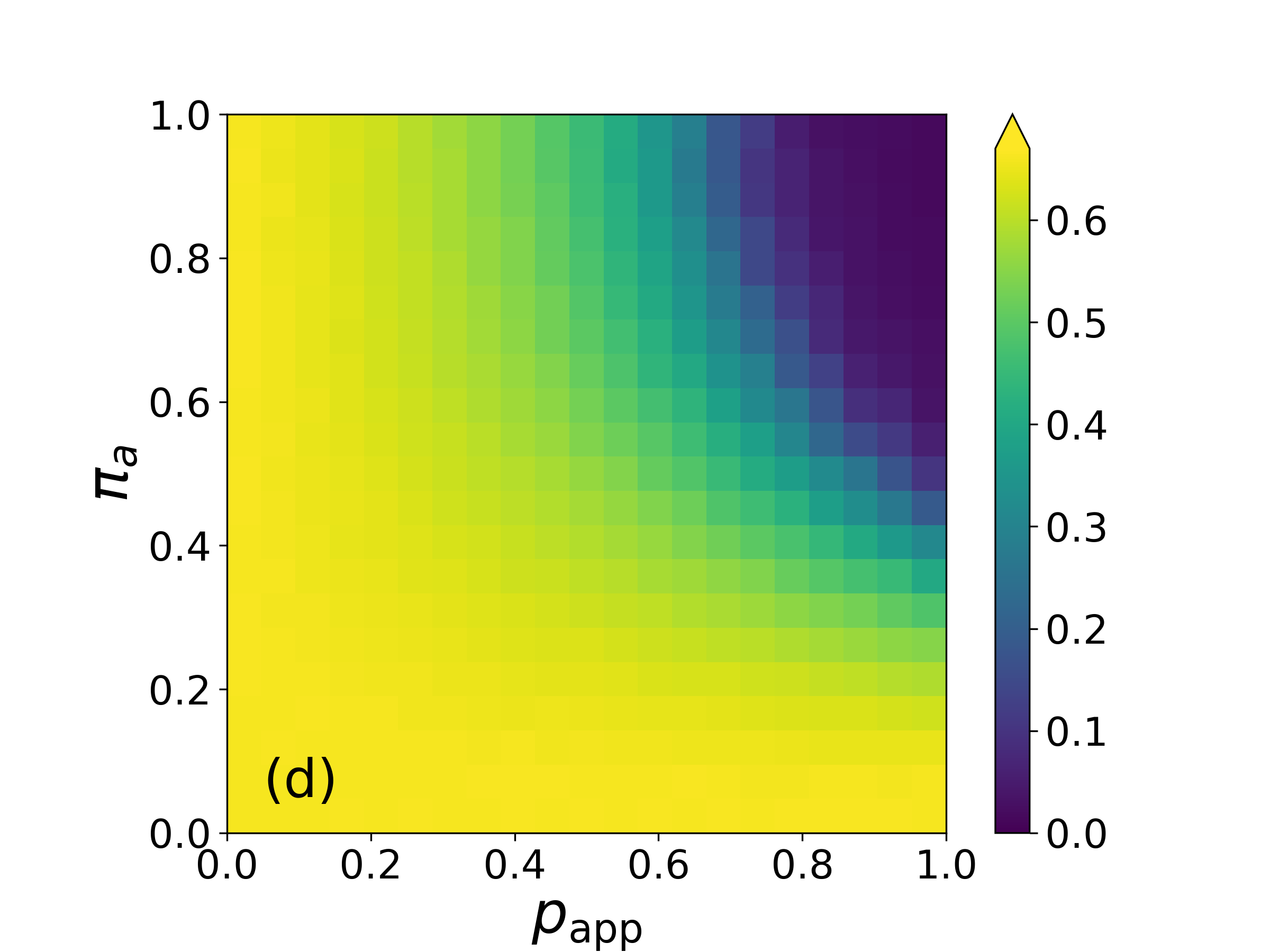}
    \caption{Expected epidemic size in the case of quarantine failures. Expected epidemic size at $\bar{k}_{\rm e} = 1.8$ for homogeneous networks with (a) random app adoption (b) and high-degree targeting strategy. Also, for heterogeneous networks with a power-law degree distribution with (c) random app adoption (d) and high-degree targeting strategy. In (b) and (d) the pattern is different due to the effects of hubs. When doing a high-degree targeting strategy, quarantine failures are more significant since the infected ones are highly influential on the spreading dynamics.}
    \label{fig:2B-appendix}
\end{figure}

\begin{figure}[H]
    \includegraphics[width=.5\linewidth]{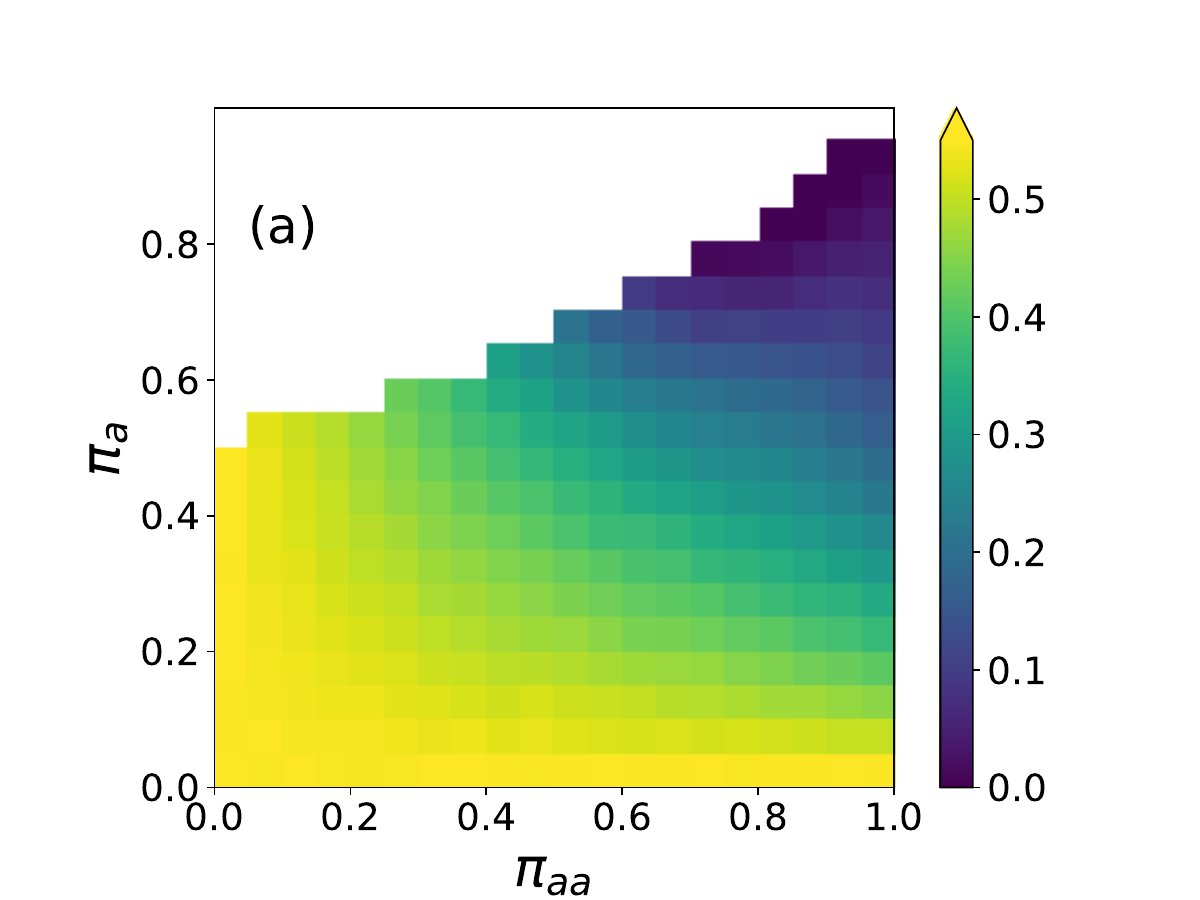}%
    \includegraphics[width=.5\linewidth]{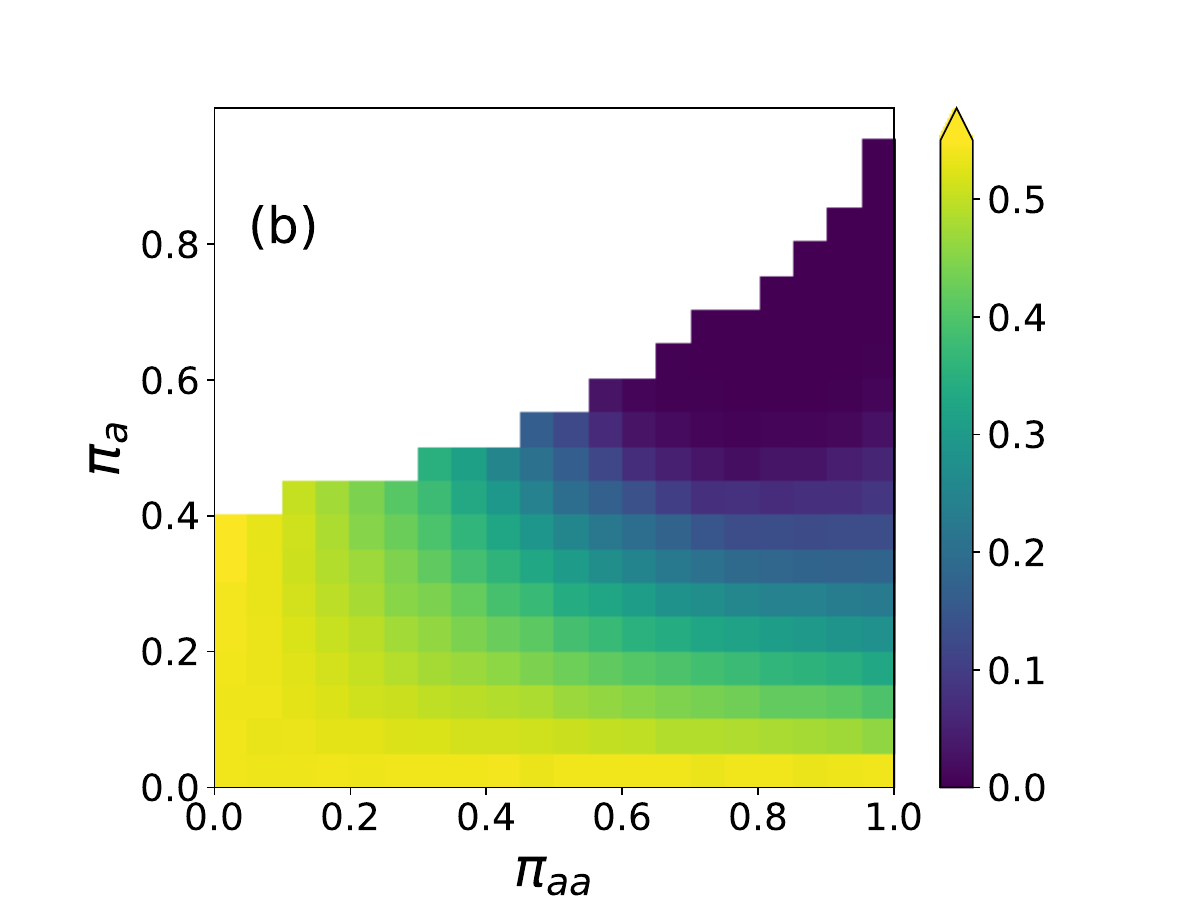}
    \includegraphics[width=.5\linewidth]{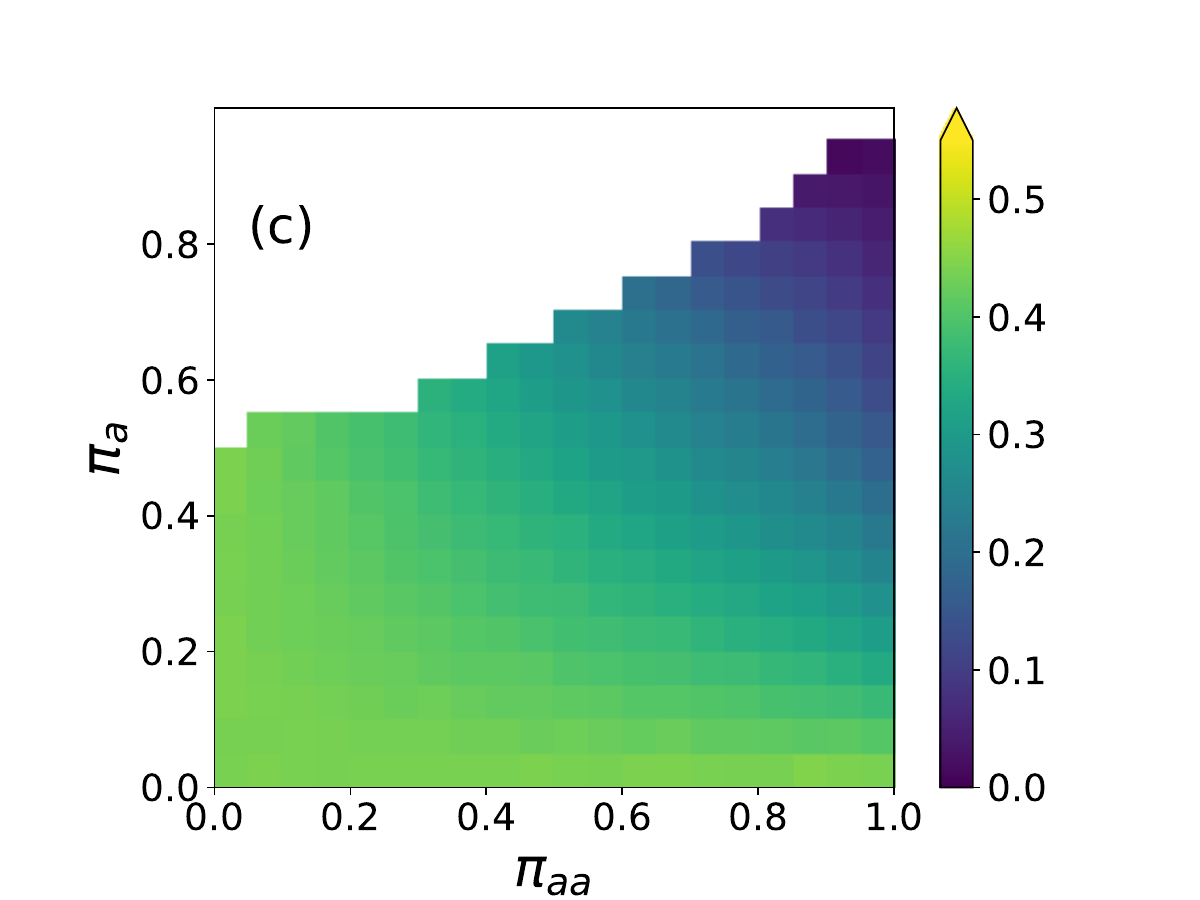}%
    \includegraphics[width=.5\linewidth]{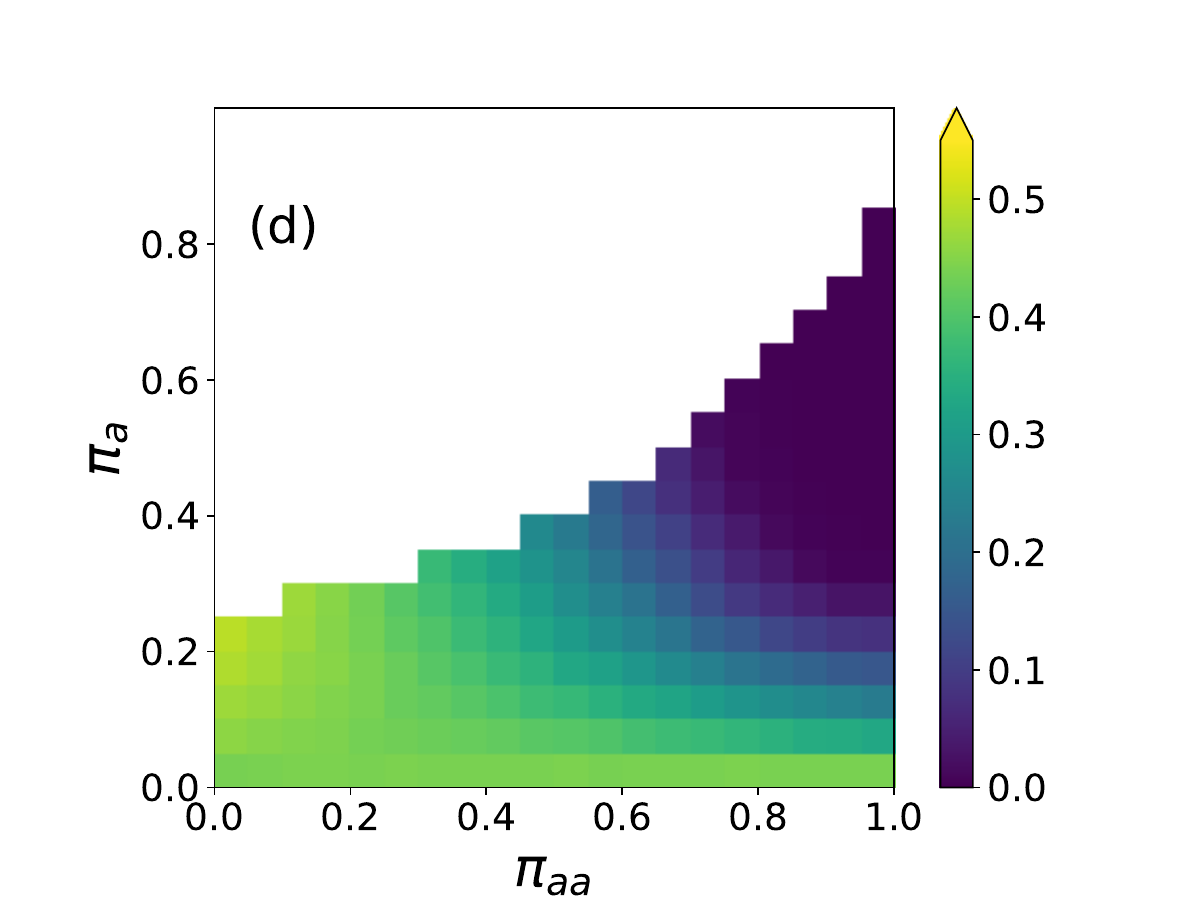} 
    \caption{The effect of homophily/heterophily in app adoption on the expected epidemic size. Expected epidemic size at $\bar{k}_{\rm e} = 1.8$ from percolation simulations for homogeneous networks with (a) random app adoption (b) and high-degree targeting strategy. Also, for heterogeneous networks with a power-law degree distribution with (c) random app adoption (d) and high-degree targeting strategy. The empty white region is the spectrum that having such a homo/heterophilic population is impossible.}
    \label{fig:4x}
\end{figure}

\begin{figure}[H]
    \includegraphics[width=.5\linewidth]{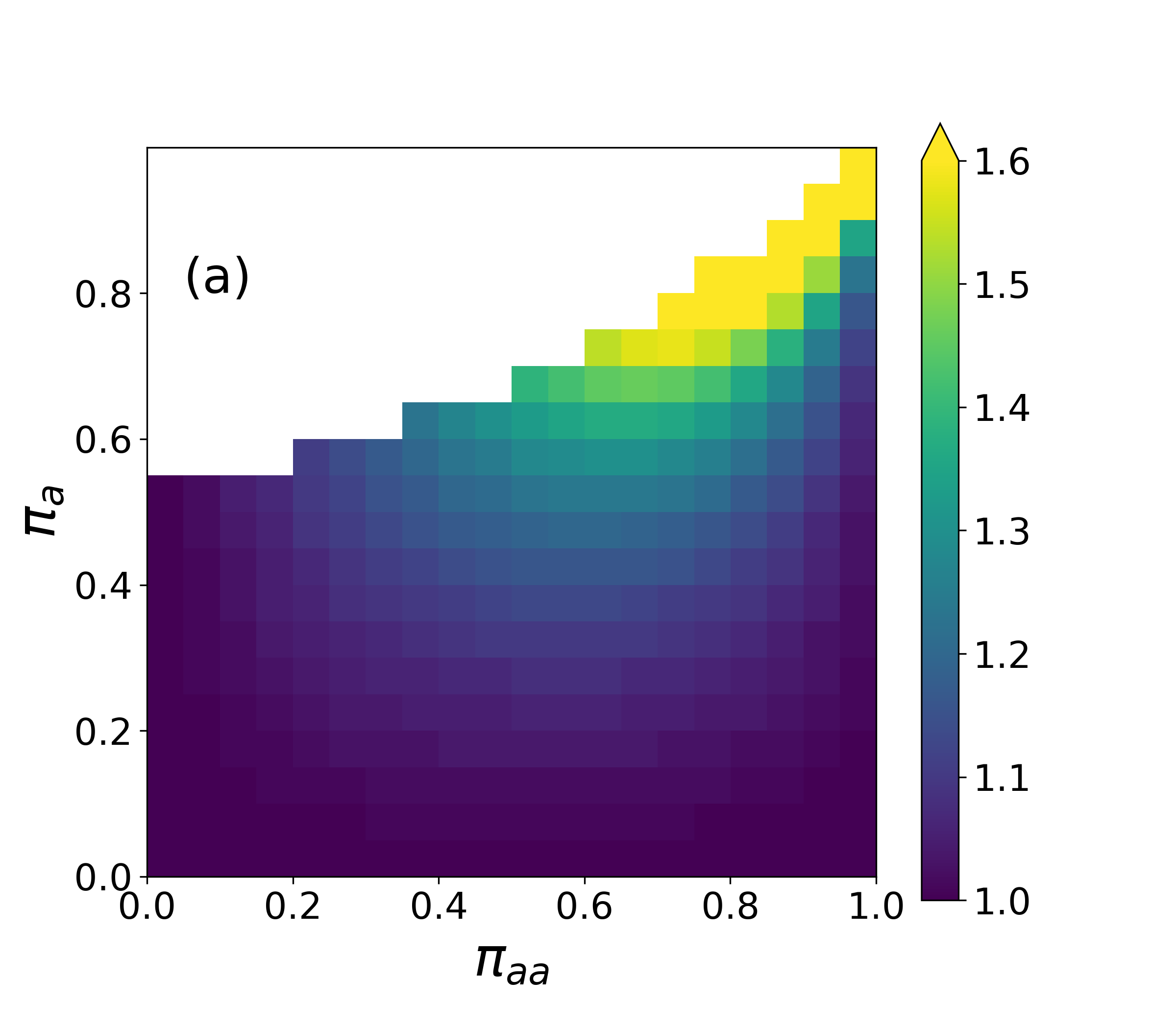}%
    \includegraphics[width=.5\linewidth]{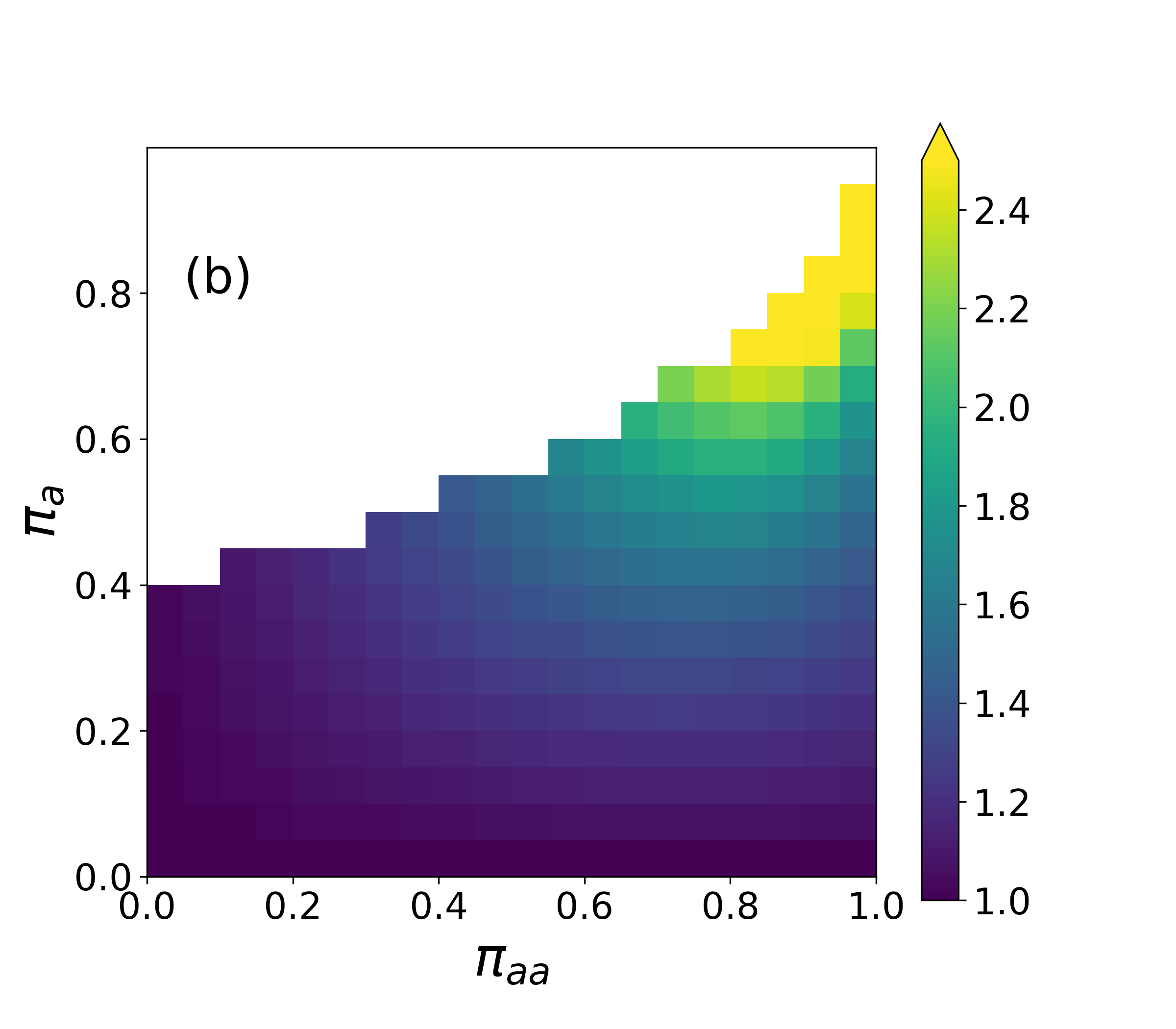}

    \includegraphics[width=.5\linewidth]{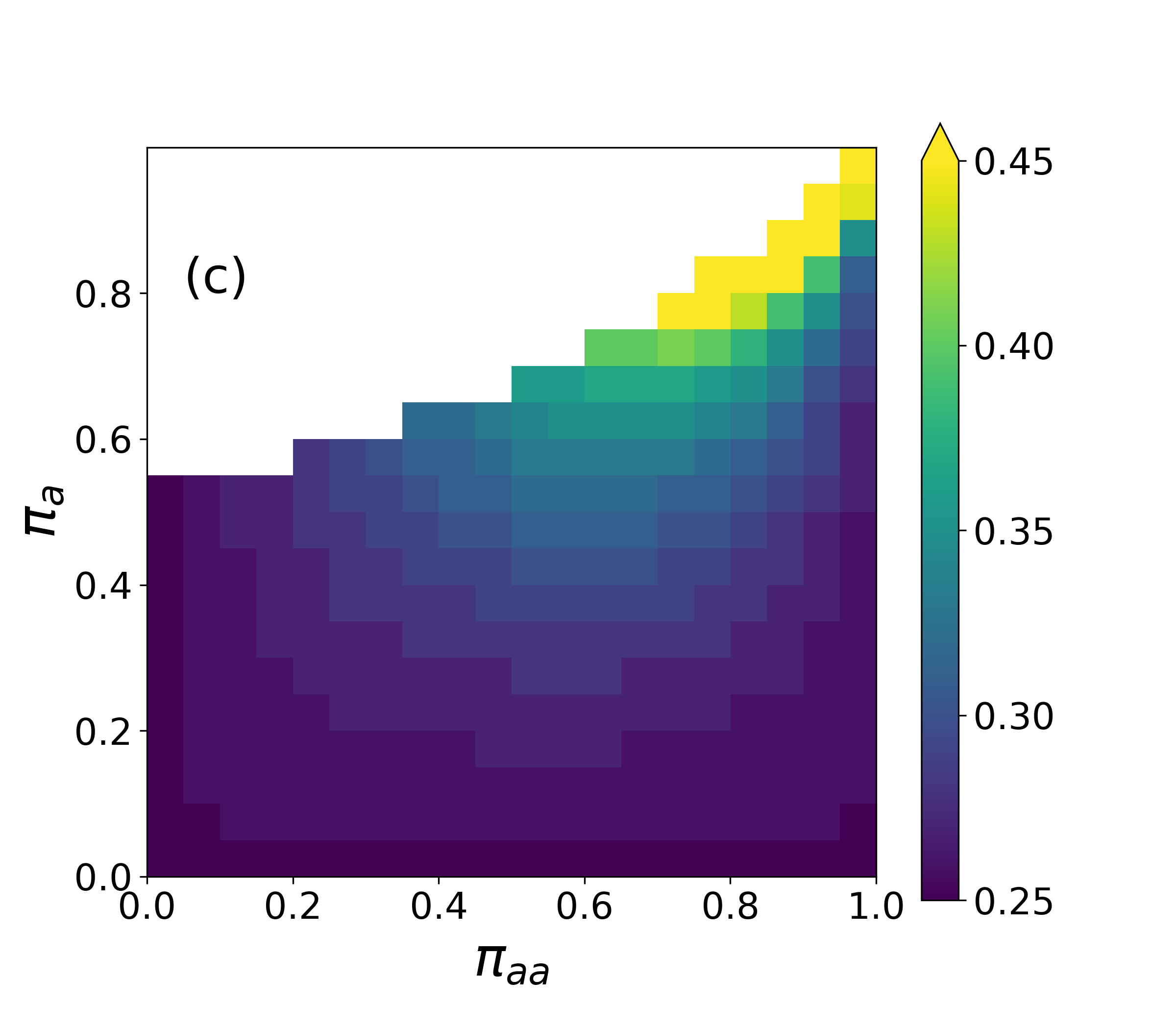}%
     \includegraphics[width=.5\linewidth]{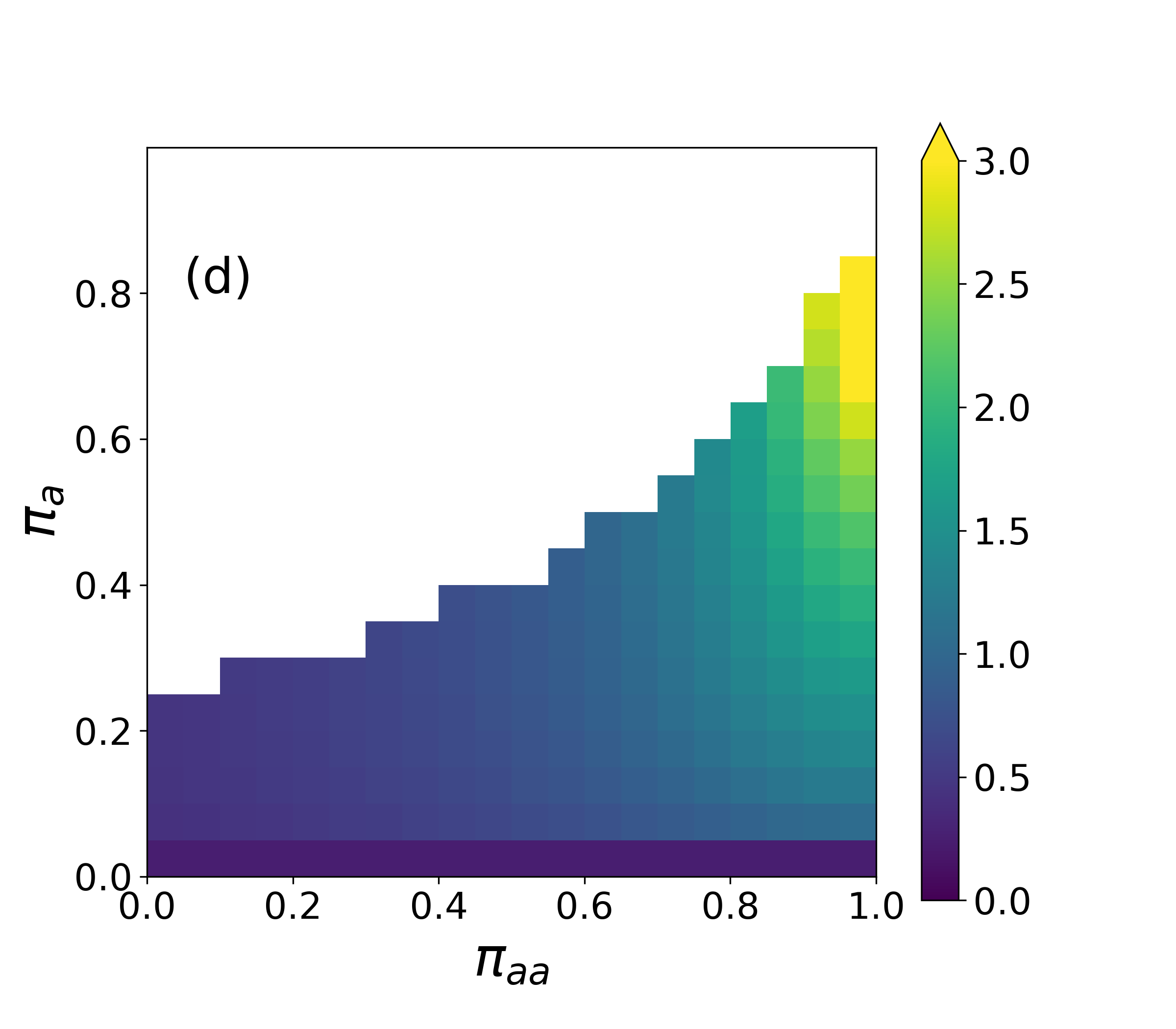}
    \caption{The effect of homophily/heterophily in app adoption on the epidemic threshold and optimum pattern for homophily. Epidemic thresholds for homogeneous networks with (a) random app adoption (b) and high-degree targeting strategy. Also, for heterogeneous networks with a power-law degree distribution with (c) random app adoption (d) and high-degree targeting strategy. The empty white region is the spectrum that having such a homo/heterophilic population is impossible.}
    \label{fig:4zz}
\end{figure}

About quarantine failures, as it was shown in Fig.~\ref{fig:2}, Fig.~\ref{fig:2A-appendix} and \ref{fig:2B-appendix} also show that contact tracing can yield very good results in terms of reducing the epidemic threshold and expected epidemic size if everything goes right at least for 50\% and half of the people use the apps. This effect is more prominent if we go for the high-degree targeting strategy, especially in heterogeneous networks, as shown in Fig.~\ref{fig:2A-appendix}d and \ref{fig:2B-appendix}d. 
Fig.~\ref{fig:4x} and \ref{fig:4zz} show that there is an optimum value for homophily in app adoption as it was shown in Fig.~\ref{fig:4} and Fig.~\ref{fig:4_branching}. The only exception is when we follow a high-degree targeting strategy in heterogeneous networks. In this case, we can see the hub effect on the epidemic threshold and size.  

\end{document}